\documentclass[11pt]{article}
\usepackage{geometry}
\usepackage{footnote}
\usepackage{booktabs,bigstrut}
\usepackage{algorithm}
\usepackage[english]{babel}
\usepackage{amsmath,amsthm,amssymb,xcolor}
\usepackage{wrapfig}
\usepackage{multicol}
\usepackage{float}
\usepackage{epsfig}
\usepackage{graphicx}
\usepackage{amsfonts}
\usepackage{bbm}
\usepackage{stmaryrd}
\usepackage{natbib}
\usepackage{xspace}
\usepackage{listings}
\usepackage{pgf} 
\usepackage{array,color,colortbl}
\usepackage{mathrsfs}
\usepackage{adjustbox}
\usepackage{dsfont}
\usepackage{hyperref}
\usepackage{algorithm}
\usepackage{algpseudocode}
\usepackage{enumitem}

\usepackage{tabularx}
\usepackage{caption}
\usepackage{subcaption}

\usepackage{mathtools}  
\mathtoolsset{showonlyrefs=true}  

\usepackage{placeins}  

\algnewcommand{\Inputs}[1]{%
  \State \textbf{Inputs:}
  \Statex \hspace*{\algorithmicindent}\parbox[t]{.8\linewidth}{\raggedright #1}
}
\algnewcommand{\Initialize}[1]{%
  \State \textbf{Initialize:}
  \Statex \hspace*{\algorithmicindent}\parbox[t]{.8\linewidth}{\raggedright #1}
}

\newtheorem{theorem}{Theorem}[section]
\newtheorem{lemma}{Lemma}
[section]
\newtheorem{proposition}{Proposition}
[section]
\newtheorem{corollary}{Corollary}
[section]
\newtheorem{conjecture}{Conjecture}
[section]

\newtheorem{remark}{Remark}[section]
\newtheorem{comment}{Comments}[section]
\newtheorem{example}{Example}[section]
\newtheorem{notation}{Notation}[section]
\newtheorem{definition}{Definition}[section]
\newtheorem{assumption}{Assumption}[section]

\newcommand{\bt}{\begin{theorem}}\newcommand{\et}{\end{theorem}} 
\newcommand{\bl}{\begin{lemma}}\newcommand{\el}{\end{lemma}}
\newcommand{\bp}{\begin{proposition}}\newcommand{\ep}{\end{proposition}}
\newcommand{\bcor}{\begin{corollary}}\newcommand{\ecor}{\end{corollary}}
\newcommand{\bconj}{\begin{conjecture}}\newcommand{\econj}{\end{conjecture}}

\newcommand{\bd}{\begin{definition} \rm}
\newcommand{\ed}{\end{definition} }
\newcommand{\brem }{\begin{remark}}
\newcommand{\erem }{\end{remark}}
\newcommand{\bcom}{\begin{comment} \rm }\newcommand{\ecom }{\end{comment}}
\newcommand{\brems }{\begin{remark} 
}
\newcommand{\erems }{\end{remark}}
\newcommand{\bex}{\begin{example} \rm}
\newcommand{\eex}{\end{example}}
\newcommand{\bno}{\begin{notation} \rm}
\newcommand{\eno}{\finproof\end{notation}}\newcommand{\enos}{\end{notation}}
\newcommand{\eexs}{\end{example} }
\newcommand{\bhyp}{\begin{assumption} \rm }\newcommand{\ehyp}{\end{assumption}}

\providecommand{\keywords}[1]{\textbf{\textit{keywords.}}#1}

\providecommand{\subjclass}[2]{\textbf{\textit{JEL Classification.}}#1}

\hypersetup{
 colorlinks  = true, %
 urlcolor   = black, %
 linkcolor  = black, 
 citecolor  = black %
}

\begin{document}
\title{Provisions  and Economic Capital for Credit Losses}
\author{{Dorinel Bastide\textsuperscript{a, b},
Stéphane Crépey\textsuperscript{c}}}
\date{{\small{}{}{}{}This version: \today}}
\maketitle

\begin{abstract}
Based on supermodularity ordering properties, we show that convex risk measures of
credit losses are nondecreasing  w.r.t.\ credit-credit and, in a wrong-way risk setup, credit-market, covariances of elliptically distributed latent factors. These results support the use of such setups for computing credit provisions and economic capital or for conducting stress test exercises and risk management analysis.
\end{abstract}

\noindent\keywords{supermodular function, convex risk measure, supermodular random variable, stop loss property,  elliptical distribution, credit loss.}

\vspace{1.5mm}
\noindent\subjclass{ G13, G17, G21.}\\


{\let\thefootnote\relax\footnotetext{\textsuperscript{a} \textit{BNP Paribas Stress Testing Methodologies \& Models. This article represents the opinions of the author, and it is not meant to represent the position or opinions of BNP Paribas or its members.} dorinel.2.bastide@bnpparibas.com  \textbf{(corresponding author).}}}

{\let\thefootnote\relax\footnotetext{\textsuperscript{b} \textit{Laboratoire de Mathématiques et Modélisation d'Evry (LaMME), Université d'Evry/Université Paris-Saclay CNRS UMR 8071.}}}

{\let\thefootnote\relax\footnotetext{\textsuperscript{c} \textit{Université Paris Cité / LPSM CNRS UMR 8001,  Paris, France}}}

\sloppy  

\section{Introduction}

Elliptical distributions are largely used in finance modelling, be it for credit latent variables or portfolio positions modelling \citep{McNeilFreyEmbrechts2015}. 
They allow to generate, by means of Monte-Carlo routines, a span of possible losses. 
Risk indicators can then be formed so as to
monitor possible future losses borne by a financial institution.
In particular, regulatory bodies instruct to rely on value-at-risk and expected shortfall measures, where the last type, which falls under the remit of coherent risk measure\footnote{see e.g.\ \citet*{AT2002}.}, is often preferred to quantiles usage. One example can be found in \citet*{BastideCrepeyDrapeauTadese21a}, where the economic capital of a clearing member bank of a central counterparty (CCP) is based on an expected shortfall risk measure of the bank loss over one year. Such measure of loss depicts numerically a nondecreasing property w.r.t.\ credit-credit and credit-market dependence parameters, capturing an increase in loss given default amounts as more defaults materialize.
\citet*{CL08} outlined the possible application of supermodular order for comparing CDO tranche premiums w.r.t.\ a credit correlation parameter of their default latent variables modelled as Brownian motions. \citet*{MS2012} generalize the use of such notion, citing application to credit losses with bounded support.
The supermodular order property finds its root in \citet*[Definition C.2, page 146]{MarshallOlkinArnold1979} under the name of lattice-superadditive property.
It has attracted subsequent attention with the works of \citet*{M97} and \citet*{BM98}, applied to stop-loss ordering of aggregated losses. \citet*{BM05} emphasize the role of several stochastic orders in relation with convex risk measures. In particular, multidimensional elliptically distributed random variables have the supermodular order property w.r.t.\ their covariance matrix coefficients (see \citet*[Corollary 2.3]{BS88} recalled in Section \ref{s:ellipt}). This result will play a key role in this paper.

In \citet*{BastideCrepeyDrapeauTadese21a}, the loss takes a more complex form than what is usually found in the credit risk literature such as \citet*{CL08}. This is due to a loss allocation coefficient attributed by the CCP to the surviving members.
In this paper, we prove the nondecreasing property of convex risk measures w.r.t.\ covariance coefficients of portfolio credit losses.
Our main motivation is to provide evidence of the soundness of the related approaches for computing credit provisions such as current expected credit loss (CECL, akin to the ${\rm CVA}$ in the central clearing one-period XVA setup of \citet*{BastideCrepeyDrapeauTadese21a}), and economic capital (EC).
This is important in justifying model assumptions and design, part of the model development cycle advocated by regulators and supervisors \citep{ECB2019}. Table \ref{tab:MetricsOfInterest} details the two main targeted metrics in this work.
{
Our main contribution is the identification of supermodular properties of the credit loss function for collateralized derivatives portfolios. Such property allows us to show the monotonous properties of convex risk measures applied to those portfolio credit losses with respect to covariance coefficients in elliptical models. We also outline such monotonic property for the contract value of credit multi-names derivatives.
}

\begin{table}[htp]
\small
\begin{center}
\caption{Metrics of interest ($.^0$ relates to the survival probability measure of the reference bank {labelled by $0$}).}
{\begin{tabular}{ ccc }
 \hline
 \textbf{Name} & \textbf{Expression} & \textbf{Reference} \\ \hline
 \begin{tabular}{c}
      current expected \\ credit loss (CECL)
 \end{tabular}
 & $\mathbb{E}^0\left(\displaystyle\sum_{i=1}^n f_i(X_1,\dots,X_n)g_i(Y_i)\right)$ & 
 Definition \ref{d:CVA_ECL_Def} \\ \hline
 \begin{tabular}{c} economic capital (EC)
 \end{tabular}
 & $\mathbb{ES}^0_\alpha\left(\displaystyle\sum_{i=1}^n f_i(X_1,\dots,X_n)g_i(Y_i)\right)$ & 
 \begin{tabular}{c}
 Definitions \ref{d:EC_KVA_Def} \\
   and \ref{d:ESdef}    
 \end{tabular}
\end{tabular}
}
\label{tab:MetricsOfInterest}
\end{center}
\end{table}

The paper is organized as follows.
{
Section \ref{s:CCR_SM} presents the arguments based on the supermodular functions and random vectors properties that will serve to prove our main result Theorem \ref{them:MonToSM}. Results on the monotonicity of credit provisions and economic capital metrics w.r.t.\ covariance coefficients in elliptical models applied to exposures towards CCP are given in Section \ref{s:CCPExpl}, with numerical illustration provided in Section \ref{s:NumericsUseCaseCCP}. Section \ref{s:CrDerivSM} completes the results from \citet*{CL08} regarding equity and senior CDO tranches. Section \ref{s:Conclusion} synthesizes our results. Supermodular functions, elliptical distributions and risk measures are reviewed in Sections \ref{apdx:SMFns}, \ref{s:ellipt} and \ref{s:RMordPpties}.

}

\subsection{Probabilistic Setup}
We consider a non-atomic
probability space $(\Omega,\mathcal{A}, \mathbb{Q})$ with a probability measure $\mathbb{Q}$ interpreted as risk-neutral in financial applications, and corresponding expectation, variance and covariance operators denoted by $\mathbb{E}$, $\mathbb{V}{\rm ar}$ and $\mathbb{C}{\rm ov}$. 


{

We consider random vectors (where all rows are considered as column vectors) $\mathbf{X}=(\mathcal{X}_1,\dots,\mathcal{X}_m)$ and $\mathbf{X}'=(\mathcal{X}'_1,\dots,\mathcal{X}'_m)$ (for some $m\in\mathbb{N}^\ast$).
The metrics of interest in this work are considered from a reference market participant viewpoint, namely a bank indexed by $0$.
In this context, it is sometimes useful to introduce a measure $\mathbb{Q}^0$ defined in terms of a measurable function $h$ of a latent variable $\mathcal{X}_0$ of the default of the bank, such that
\begin{align}\label{e:MeasChgSpec}
h(\mathcal{X}_0)=d\mathbb{Q}^0/d\mathbb{Q}\geq 0\,\,\mbox{ and }\,\, \mathbb{E}\left[h(\mathcal{X}_0)\right]=1.
\end{align}
}

{
\begin{example}
In the setup of \citet*{BastideCrepeyDrapeauTadese21a}, even though the financial risk factors and the financial metrics are defined under $\mathbb{Q}$, more explicit XVA formulas arise in terms of the related bank survival probability measure $\mathbb{Q}^0$.
\end{example}
\noindent We index all the formerly introduced probabilistic notations by ``$.\,{^0}\,$'' whenever applied in reference to $\mathbb{Q}^0$.
}

\begin{assumption}\label{a:HypCondSM}
{$\mathcal{X}_0$,$\mathbf{X}$ and $\mathbf{X}'$}
satisfy
\begin{align}\label{e:hypSMcondX0equiv}
\big[{\mathbf{X}}\big|\mathcal{X}_0\big]\leq_{{sm}}\big[\mathbf{X}'\big|\mathcal{X}_0\big],
\end{align}
i.e.
\begin{align}\label{e:hypSMcondX0}
\mathbb{E}\left[f(\mathbf{X})\big|\mathcal{X}_0\right]\leq\mathbb{E}\left[f(\mathbf{X}')\big|\mathcal{X}_0\right]
\end{align}
holds for any supermodular function $f:\mathbb{R}^{m}\longrightarrow\mathbb{R}$ such that the conditional expectations exist.
\end{assumption}



\begin{lemma}\label{l:SurvMeasSMPreserv}
If $\mathcal{X}_0$, $\mathbf{X}$ and $\mathbf{X}'$ satisfy Assumption \ref{a:HypCondSM}, then $\mathbf{X}\leq_{sm^0}\mathbf{X}'$.
\end{lemma}
\proof
As $h(\mathcal{X}_0)\geq 0$, for any supermodular function $f$ on $\mathbb{R}^{m}$ such that both $\mathbb{E}^0\left[f(\mathbf{X})\right]$ and $\mathbb{E}^0\left[f(\mathbf{X}')\right]$ exist, \eqref{e:MeasChgSpec} and \eqref{e:hypSMcondX0} yield
\begin{align}
\begin{split}
    \mathbb{E}^0\left[f(\mathbf{X})\right] & = \mathbb{E}\left[h(\mathcal{X}_0)f(\mathbf{X})\right]
     = \mathbb{E}\left[h(\mathcal{X}_0)\mathbb{E}\left[f(\mathbf{X})\big|\mathcal{X}_0\right]\right]\\
     & \leq \mathbb{E}\left[h(\mathcal{X}_0)\mathbb{E}\left[f(\mathbf{X}')\big|\mathcal{X}_0\right]\right]
     = \mathbb{E}^0\left[f(\mathbf{X}')\right].~\square
\end{split}
\end{align}

For a function of several real arguments,
by nondecreasing we mean nondecreasing w.r.t.\ each of them.
\begin{lemma}\label{l:fromSMtoSL}
If $\mathbf{X}\leq_{sm^0}\mathbf{Y}$, then, for any nondecreasing supermodular function $f:\mathbb{R}^m\rightarrow\mathbb{R}$, $f(\mathbf{X})\leq_{sl^{0}}f(\mathbf{Y})$ where $\leq_{sl^0}$ denotes the stop-loss order between random vectors under $\mathbb{Q}^0$ as per \citet[Definition 2.1]{M97} recalled in Appendix \ref{apdx:SMFns}.
\end{lemma}
\proof
For any $A\in\mathbb{R}$, the function $\varphi:\mathbb{R}\rightarrow\mathbb{R}_+$, $ x\mapsto(x-A)^+$ is nondecreasing and convex. By \citet[Theorem 3.9.3 f), page 113]{MS2002}, $\varphi\circ f$ is nondecreasing supermodular. Hence
$\mathbb{E}^{0}\big[(f(\mathbf{X})-A)^+\big]\leq\mathbb{E}^{0}\big[(f(\mathbf{Y})-A)^+\big]$, which yields the result by \citet[Definition 2.1]{M97}.~$\square$

\paragraph{Elliptical setup} Whenever stated in the paper, by elliptical setup,
we mean that $(\mathcal{X}_0,\mathbf{X})$ and $(\mathcal{X}_0,\mathbf{X}')$ follow elliptical distributions (see Section \ref{s:ellipt}) under $\mathbb{Q}$ as per
\begin{align}\label{e:ModelElliptic}
(\mathcal{X}_0,\mathbf{X})={\boldsymbol \mu}
+\mathbf{A}\mathbf{Z}\mbox{ and }(\mathcal{X}_0,\mathbf{X}')={\boldsymbol \mu}+\mathbf{A}'\mathbf{Z},
\end{align}
for constant matrices $\mathbf{A}$, $\mathbf{A}'\in\mathbb{R}^{m\times k}$ of full rank with the same first row. $\mathbf{Z}=(\mathcal{Z}_0,\mathcal{Z}_1,\dots,\mathcal{Z}_{k})$ follows a spherical distribution $S_{k+1}(\psi)$ (see Section \ref{s:ellipt}), with characteristic generator $\psi$. The rationale for having the same $\mathcal{X}_0$ in $(\mathcal{X}_0,\mathbf{X})$ and $(\mathcal{X}_0,\mathbf{X}')$ is that the point of view will be the one of a reference bank indexed by $0$
(cf. Lemma \ref{lem:SpecCaseElliptic} and Proposition \ref{p:ConclResFmLemmas}).
\noindent We denote by
$\boldsymbol\Gamma=\mathbf{A}\mathbf{A}^\top$ and $\boldsymbol\Gamma'=\mathbf{A}'(\mathbf{A}')^\top$ the $\mathbb{Q}$ covariance matrices of $(\mathcal{X}_0,\mathbf{X})$ and $(\mathcal{X}_0,\mathbf{X}')$, assumed to be positive semi-definite. We  write $\Gamma_{ij}=\mathbb{C}{\rm ov}(\mathcal{X}_i,\mathcal{X}_j)$ for all $i, j\in 0\, ..\, m$, and likewise for $\boldsymbol\Gamma'$.
Moreover, we assume within this setup that $\Gamma_{jj}=\Gamma'_{jj}$ and $\Gamma_{0j}=\Gamma'_{0j}, j\in 1\, ..\, m,$ and $\Gamma'_{ij}\leq\Gamma'_{ij}, i\neq j\in 1\, ..\, m$.
\noindent In particular, $\mathbf{X}\leq_{sm
}\mathbf{X}'$, by \citet*[Corollary 2.3]{BS88} recalled in Section \ref{s:ellipt}. 

\begin{example}
By Proposition \ref{p:ConclResFmLemmas}, the elliptical setup \eqref{e:ModelElliptic} verifies Assumption \ref{a:HypCondSM}.
\end{example}

\section{Theoretical Results}\label{s:CCR_SM}


Given measurable functions $f_i:\mathbb{R}^n\rightarrow\mathbb{R}_{+}$ and $g_i:\mathbb{R}\rightarrow\mathbb{R}_{+}$, $i\in 1\, ..\, n$, let
\begin{equation}\label{e:LossForm}
\begin{array}{cccl}
     & \mathbb{R}^{2n} & \overset{\ell}{\longrightarrow} & \mathbb{R}_+ \\
     & (x_1,\dots,x_n,y_1,\dots,y_n) & \longmapsto & \displaystyle\sum_{i=1}^n f_i(x_1,\dots,x_n)g_i(y_i). 
\end{array}
\end{equation}
We additionally assume that all the functions $f_i$ are supermodular and nondecreasing on $\mathbb{R}^n$ and all the functions $g_i$ are nondecreasing on $\mathbb{R}$.
\begin{lemma}\label{l:MultCrossVbSM}
$\ell$ has increasing differences w.r.t.\ any pair $(x_i,y_j)\in\mathbb{R}^{2}$, $i,j\in 1\, ..\, n$.
\end{lemma}
\proof Let $f_i(x_j|\mathbf{x}_{-j})$ denote the function $f_i$ applied to $x_j$ but keeping all other arguments $\mathbf{x}_{-j}:=\left(x_k\right)_{k\neq j}$ fixed. We look at the two cases where we consider either a pair of argument $(x_i,y_i)\in\mathbb{R}^2$, $i\in 1\, ..\, n$ (i.e.\ the pair of arguments tested for the increasing difference are part of the same term of the sum), or a pair $(x_i,y_j)\in\mathbb{R}^2$, $i\neq j\in 1\, ..\, n$, and the corresponding increasing differences.\\
\textit{Case $(x_i,y_i)\in\mathbb{R}^2$, $i\in 1\, ..\, n$:} the function $\mathbb{R}^2\ni (x_i,y_i)\mapsto f_i(x_i|\mathbf{x}_{-i})g_i(y_i)$, has the increasing differences property by application of Lemma \ref{l:crossVbSM} with $g(\cdot)=f_i(\cdot|\mathbf{x}_{-i})$ and $h(\cdot)=g_i(\cdot)$. This increasing difference property writes
\begin{align*}
& f_i(x_i'|\mathbf{x}_{-i})g_i(y_i')+\sum_{k\neq i}f_k(x'_i|\mathbf{x}_{-i})g_k(y_k)-f_i(x_i'|\mathbf{x}_{-i})g_i(y_i)-\sum_{k\neq i}f_k(x'_i|\mathbf{x}_{-i})g_k(y_k)\\
&- f_i(x_i|\mathbf{x}_{-i})g_i(y_i')-\sum_{k\neq i}f_k(x_i|\mathbf{x}_{-i})g_k(y_k)+f_i(x_i|\mathbf{x}_{-i})g_i(y_i)+\sum_{k\neq i}f_k(x_i|\mathbf{x}_{-i})g_k(y_k)\\
= & \underbrace{f_i(x_i'|\mathbf{x}_{-i})g_i(y_i')-f_i(x_i'|\mathbf{x}_{-i})g_i(y_i)-f_i(x_i|\mathbf{x}_{-i})g_i(y_i')+f_i(x_i|\mathbf{x}_{-i})g_i(y_i)}_{\geq 0\mbox{ \scriptsize by Lemma \ref{l:crossVbSM}}}\\
&\underbrace{+\sum_{k\neq i}f_k(x'_i|\mathbf{x}_{-i})g_k(y_k)-\sum_{k\neq i}f_k(x'_i|\mathbf{x}_{-i})g_k(y_k)}_{=0}\\
& \underbrace{-\sum_{k\neq i}f_k(x_i|\mathbf{x}_{-i})g_k(y_k)+\sum_{k\neq i}f_k(x_i|\mathbf{x}_{-i})g_k(y_k)}_{=0}\\
\geq & 0,
\end{align*}
hence $(x_i,y_i)\mapsto \sum_{k=1}^n f_k(x_i|\mathbf{x}_{-i})g_k(y_k)$ has the increasing differences property.\smallskip

\noindent\textit{Case $(x_i,y_j)\in\mathbb{R}^2$, $i\neq j\in 1\, ..\, n$:} we write the increasing difference
\begin{align*}
& \sum_{k\neq j} f_k(x'_i|\mathbf{x}_{-i})g_k(y_k)+f_j(x'_i|\mathbf{x}_{-i})g_j(y'_j)-\sum_{k\neq j} f_k(x'_i|\mathbf{x}_{-i})g_k(y_k)-f_j(x'_i|\mathbf{x}_{-i})g_j(y_j)\\
- & \sum_{k\neq j} f_k(x_i|\mathbf{x}_{-i})g_k(y_k)-f_j(x_i|\mathbf{x}_{-i})g_j(y'_j)+
\sum_{k\neq j} f_k(x_i|\mathbf{x}_{-i})g_k(y_k)+f_j(x_i|\mathbf{x}_{-i})g_j(y_j)\\
= & f_j(x'_i|\mathbf{x}_{-i})g_j(y'_j) -f_j(x'_i|\mathbf{x}_{-i})g_j(y_j)-f_j(x_i|\mathbf{x}_{-i})g_j(y'_j)+f_j(x_i|\mathbf{x}_{-i})g_j(y_j)\geq 0,
\end{align*}
by application of Lemma \ref{l:crossVbSM} with $g(\cdot)=f_j(\cdot|\mathbf{x}_{-i})$ and $h(\cdot)=g_j(\cdot)$.$\square$
\begin{proposition}\label{p:MultCrossVbSMbilat}
In the special case $f_i(x_1,\dots,x_n)\equiv f_i(x_i)$ with $f_i$ nondecreasing on $\mathbb{R}$, $i\in 1\,..\,n,$ we have that
$\mathbb{R}^{2n}\ni(x_1,\dots,x_n,y_1,\dots,y_n)\overset{\ell}{\mapsto} \sum_{i=1}^n f_i(x_i)g_i(y_i)\in\mathbb{R}$ has increasing differences w.r.t.\ any pair $(x_i,y_j)\in\mathbb{R}^{2}$, $i,j\in 1\, ..\, n$.
\end{proposition}
\proof
By Lemma \ref{lem:aggLossRkSM}, $\ell$ has increasing differences w.r.t.\ any pair $(x_i,x_j)\in\mathbb{R}^{2}$, $i\neq j\in 1\, ..\, n$, as well as any pair $(y_i,y_j)\in\mathbb{R}^{2}$, $i\neq j$. For $i\in 1\, ..\, n$, $(x_i,y_i)\mapsto f_i(x_i)g_i(y_i)$, has the increasing differences property, by application of Lemma \ref{l:crossVbSM} with $g(\cdot)=f_i(\cdot)$ and $h(\cdot)=g_i(\cdot)$. Hence $\ell$ has the increasing differences property w.r.t.\ any pair $(x_i,y_i)\in\mathbb{R}^2$. Finally, for $i\neq j\in 1\, ..\, n$, $(x_i,y_j)\mapsto f_i(x_i)g_i(y_i)+f_i(x_j)g_j(y_j)$ has the increasing differences property by Lemma \ref{lem:aggLossRkSM} with $h_i(\cdot)=f_i(\cdot)g_i(y_i)$ and $h_j(\cdot)=f_j(x_j)g_j(\cdot)$.
~$\square$
\begin{proposition}\label{p:GenCaseSM}
The function $\ell$
is nondecreasing supermodular.
\end{proposition}
\proof
By Lemma \ref{lem:aggLossRkSM} with $h_i(\cdot)=f_i(x_1,\dots,x_n)g_i(\cdot)$, for any $i\in 1\, ..\, n$, $\ell$ has increasing differences with respect to any pair $(y_k,y_l)$, $k,l\in 1\, ..\, n$. By assumption and closure by addition of the increasing differences property, $\ell$ has increasing differences with respect to any pair $(x_k,x_l)$, $k,l\in 1\, ..\, n$. Finally, by Lemma \ref{l:MultCrossVbSM}, $\ell$ has increasing differences with respect to any pair $(x_k,y_l)$, $k,l\in 1\, ..\, n$. Hence, by \citet[Corollary 1]{Yildiz2010}\footnote{see Section \ref{apdx:SMFns}.}, $\ell$ is supermodular.~$\square$\smallskip



Hereafter we consider a one-period financial market model on $(\Omega,\mathcal{A})$, assumed arbitrage-free\footnote{see for instance \citep*[Part I]{follmerSchied2016}.} with risk-neutral measure $\mathbb{Q}$. We set $(\mathcal{X}_0,\mathbf{X})=(X_0,X_1,\dots,X_n,Y_1,\dots,Y_n)$, with $\mathcal{X}_0$ re-noted $X_0$, where $Y_i$
drives a loss that obligor $i\in 1\, ..\, n$ generates if it defaults, with default of each credit name $j\in 0\, ..\, n$ (including the reference bank $0$) driven by a latent variable $X_j$ (for some $n>0$).
The variables $X_1,\dots,X_n$ represent latent variables breaching a certain threshold for translating the defaults of $n$ obligors also indexed by $1,\dots,n$.
The r.v. $f_i(X_1,\dots,X_n)$ represents a loss allocation coefficient depending on the possible defaults of obligors $1,\dots,n$ while $g_i(\mathcal{Y}_i)$ represents a nonnegative loss given default of obligor $i$. When $f_i(X_1,\dots,X_n)\equiv f_i(X_i)$ as in \citet*{CL08} or \citet*{MS2013}, $f_i(X_i)g_i(Y_i)$ can represent the loss related to a bilateral counterparty position, i.e.\ a portfolio position between the reference bank $0$ and its client $i$. As detailed in Section \ref{s:CCPExpl},
the more general case where $f_i(X_1,\dots,X_n)$ depends on several $X_j$ {is needed to} encompass the financial losses generated by clearing exposures towards a {CCP}.
This is due to the loss allocation mechanism between surviving members \citep*[{Section 3}]{BastideCrepeyDrapeauTadese21a}. It also covers the case of financial resolution funds \citep*{SRB2021}.\smallskip

\begin{theorem}\label{them:MonToSM}
Let $\mathfrak{X}^0$ be a closed linear subspace and sublattice of $L^1(\mathbb{Q}^0)$, that includes the
constants.
If $\mathbf{X}\leq_{sm^0}\mathbf{X}'$, then, for any risk measure\footnote{see Definition \ref{d:ConvexRiskMeas}.} $\rho$ on $\mathfrak{X}^0\ni\ell(\mathbf{X}),\ell(\mathbf{X}')$, we have $\rho\left(\ell(\mathbf{X})\right)\leq \rho\left(\ell(\mathbf{X}')\right)$.
In particular, {if} $\ell(\mathbf{X})$ and $\ell(\mathbf{X}')$ are $\mathbb{Q}^0$ integrable, then $\mathbb{E}^0\left[\ell(\mathbf{X})\right]\leq\mathbb{E}^0\left[\ell(\mathbf{X}') \right]$.
\end{theorem}

\proof
By Proposition \ref{p:GenCaseSM}, the function $\ell$ is supermodular on $\mathbb{R}^{2n}$. Hence, by Lemma \ref{l:fromSMtoSL}, if 
$\mathbf{X}\leq_{sm^0}\mathbf{X}'$, then
\begin{align}\label{e:TargetRel}
\ell(\mathbf{X})=\sum_{i=1}^n f_i(X_1,\dots,X_n)g_i(Y_i)\leq_{sl^0}\sum_{i=1}^n f_i(X'_1,\dots,X'_n)g_i(Y'_i)=\ell(\mathbf{X}'),
\end{align}
Having assumed the probability space non-atomic\footnote{see \citet*[Example 3.1]{BM05} for a counter-example to the monotonicity property in the case of a probability space with atom.}, \citet*[Theorem 4.4]{BM05}, given at the end of Appendix \ref{s:RMordPpties}, yields the result.
~$\square$\smallskip



\begin{remark}\label{rem:GenResAppEllipt} By Lemma \ref{l:SurvMeasSMPreserv}, the assumption $\mathbf{X}\leq_{sm^0}\mathbf{X}'$ in Theorem \ref{them:MonToSM} holds, in particular, for $(\mathcal{X}_0,\mathbf{X})$ and $(\mathcal{X}_0,\mathbf{X}')$ satisfying Assumption \ref{a:HypCondSM}.
\end{remark}

{

\begin{definition}\label{d:CVA_ECL_Def} The current expected credit loss of the reference bank $0$ is 
${\rm CECL}(\mathcal{X}_0,\mathbf{X})=\mathbb{E}^0\left[\ell(\mathbf{X})\right]$.
\end{definition}


\begin{proposition}\label{p:eclccva}
If $\mathcal{X}_0$, $\mathbf{X}$ and $\mathbf{X}'$ satisfy Assumption \ref{a:HypCondSM}, then ${\rm CECL}(\mathcal{X}_0,\mathbf{X})\leq {\rm CECL}(\mathcal{X}_0,\mathbf{X}')$ holds whenever $\ell(\mathbf{X}),\ell(\mathbf{X}')\in L^1(\mathbb{Q}^0)$.
\end{proposition}
}
\proof
By definition of ${\rm CECL}(\mathcal{X}_0,\mathbf{X})$ and application of Theorem \ref{them:MonToSM} and Remark \ref{rem:GenResAppEllipt} to $\rho\equiv \mathbb{E}^0$.~$\square$

\begin{definition}\label{d:EC_KVA_Def} The economic capital of the reference bank $0$ is ${\rm EC}(\mathcal{X}_0,\mathbf{X})=\mathbb{ES}^0_{\alpha}\big(\ell(\mathbf{X})\big)$, with $\mathbb{ES}^0_{\alpha}$ as per Definition \ref{d:ESdef} (applied under $\mathbb{Q}^0$).
\end{definition}

\begin{proposition}\label{p:eckva} If $\mathcal{X}_0$, $\mathbf{X}$ and $\mathbf{X}'$ satisfy Assumption \ref{a:HypCondSM}, then ${\rm EC}(\mathcal{X}_0,\mathbf{X})\leq {\rm EC}(\mathcal{X}_0,\mathbf{X}')$ holds whenever $\ell(\mathbf{X}),\ell(\mathbf{X}')\in L^1(\mathbb{Q}^0)$.
\end{proposition}
\proof By definition of ${\rm EC}(\mathcal{X}_0,\mathbf{X})$ and application of Theorem \ref{them:MonToSM} to $\rho\equiv\mathbb{ES}^0_{\alpha}$ using Remark \ref{rem:rhoAsES}.~$\square$


\begin{remark}\label{r:OntheNotations}
Wrong-way risk is the potential increase of the exposure a financial actor w.r.t.\ certain counterparties when their probability of default increase. A risk model should include a wrong-way risk feature in order to ensure conservative treatment. See \citet[Section 8.6.5]{gregory2015xva} for more detailed explanations and \citet*{Brigo2013} for various examples of asset classes models incorporating the wrong-way risk feature. Under the elliptical model \eqref{e:ModelElliptic} and the credit loss form \eqref{e:LossForm}, wrong-way risk holds, in the sense that both expectations and expected shortfall measures increase when correlation between credit and portfolio exposure increases, provided that an increase of the covariance between the default latent variable $X_i$ and the potential loss driver $Y_i$ leads to an increase of the loss amount $g_i(Y_i)$.
{As shown by Propositions \ref{p:eclccva} and \ref{p:eckva} with $\mathbb{C}{\rm ov}^0(X_i,Y_i)\leq \mathbb{C}{\rm ov}^0(X'_i,Y'_i)$ thanks to \eqref{e:SpecCaseElliptic} in Lemma \ref{lem:SpecCaseElliptic} applied to the elliptical setup \eqref{e:ModelElliptic}, this is indeed the case when $f$ and $g$ are nondecreasing (as assumed) in each of their arguments.}
\end{remark}

\section{Application to CCPs}\label{s:CCPExpl}
Given real constants $\beta_1,\dots,\beta_n\geq 0$, $m_1,\dots,m_n$ and $B_1,\dots,B_n$, we consider in \eqref{e:LossForm}
\begin{align}\label{e:LossFormCleargUncond}
\begin{array}{ccl}
     \mathbb{R}^n & \overset{f_i}{\longrightarrow} & \mathbb{R}_+ \\
     (x_1,\dots,x_n) & \longmapsto & \displaystyle\frac{\mathds{1}_{\{x_i\geq B_i\}}}{1+\sum_{j=1}^n \beta_j\mathds{1}_{\{x_j<B_j\}}} 
\end{array}\quad\mbox{and}\quad
\begin{array}{ccl}
     \mathbb{R} & \overset{g_i}{\longrightarrow} & \mathbb{R}_+ \\
     y_i & \longmapsto & \big(y_i-m_i\big)^+ 
\end{array},\quad i\in 1\, ..\, n.
\end{align}
Likewise, given constants $B_i$, $i\in 1\,..\,n$, $\{\tau_i\leq T\}:=\{X_i\geq B_i\}$ models the default event of participant $i\in 1\, ..\, n$. The weights $\left(1+\sum_{j=1}^n \beta_j\mathds{1}_{\{X_j<B_j\}}\right)^{-1}$ represent a stylised specification of a default fund allocation in a {CCP} setup.
Modeling the default of the reference bank $0$ by $\{X_0\geq B_0\}$ for some constant $B_0$, we now take in \eqref{e:MeasChgSpec}
\begin{align}\label{e:mch}h(x_0)=\displaystyle\frac{\mathds{1}_{\{
x_0<B_0\}}}{1-\gamma},\,\mbox{ where }\gamma=\mathbb{Q}(X_0\geq B_0),
\end{align}
so that $\mathbb{Q}^0$ is the survival measure of the reference bank associated with $\mathbb{Q}$ (see e.g.\ \citet*[Section 3]{CrepeyHoskinsonSaadeddine2019}).

The r.v.\ $Y_i={\rm  nom}_i\sigma_i G_i$, with ${\rm  nom}_i\in\mathbb{R}$, $\sigma_i>0$ constant and $G_i$ spherical (hence $\mathbb{V}{\rm ar}(Y_i)={\rm  nom}_i^2\sigma_i^2\mathbb{V}{\rm ar}(G_i)$) represents the loss of the market participant $0$ in case of the default of obligor $i$, collateralized by a corresponding amount $m_i$.


\begin{proposition}\label{pro:CrLattVbSMpart}
The function $f_i$
in \eqref{e:LossFormCleargUncond} is nondecreasing supermodular on $\mathbb{R}^n$.
\end{proposition}

\proof
{Given} \citet[Corollary 1]{Yildiz2010}, it is sufficient to show that $f_i$ has increasing differences. Let $k,l\in 1,..,n$.\\
\textit{Case $k,l\neq i$:}
Let $\mathbb{R}^2\ni(x_k,x_l)\overset{g}{\longmapsto} \frac{\mathds{1}_{\{x_i\geq B_i\}}}{\Lambda+\beta_k\mathds{1}_{\{x_k<B_k\}}+\beta_l\mathds{1}_{\{x_l<B_l\}}}\in\mathbb{R}_+$, with $\Lambda=\sum_{j\neq  k,l}\beta_j\mathds{1}_{\{x_j<B_j\}}$. For $x_k'\geq x_k$, $x_l'\geq x_l$, we form the increasing difference
\begin{align}\label{e:IncDiffCaseI}
\begin{split}
& g(x_k',x_l')-g(x_k',x_l)-g(x_k,x_l')+g(x_k,x_l) \\
&  \quad\quad= \displaystyle\frac{\mathds{1}_{\{x_i\geq B_i\}}\beta_l\big(\mathds{1}_{\{x_l<B_l\}}-\mathds{1}_{\{x_l'<B_l\}}\big)}{\mathrm{denom}_1}-\displaystyle\frac{\mathds{1}_{\{x_i\geq B_i\}}\beta_l\big(\mathds{1}_{\{x_l<B_l\}}-\mathds{1}_{\{x_l'<B_l\}}\big)}{\mathrm{denom}_2},
\end{split}
\end{align}
with $\mathrm{denom}_1=\big(\Lambda+\beta_k\mathds{1}_{\{x_k'<B_k\}}+\beta_l\mathds{1}_{\{x_l'<B_l\}}\big)\big(\Lambda+\beta_k\mathds{1}_{\{x_k'<B_k\}}+\beta_l\mathds{1}_{\{x_l<B_l\}}\big)$ and $\mathrm{denom}_2=\big(\Lambda+\beta_k\mathds{1}_{\{x_k<B_k\}}+\beta_l\mathds{1}_{\{x_l'<B_l\}}\big)\big(\Lambda+\beta_k\mathds{1}_{\{x_k<B_k\}}+\beta_l\mathds{1}_{\{x_l<B_l\}}\big)$. If $x_l\leq x_l'<B_l$ or $B_l\leq x_l\leq x_l'$, then the increasing difference \eqref{e:IncDiffCaseI} is zero as the numerators of both terms are zero. If $x_l<B_l\leq x_l'$, then both numerators in \eqref{e:IncDiffCaseI} equal $\mathds{1}_{\{x_i\geq B_i\}}\beta_l$. In this case: (i) if $x_k\leq x_k'<B_k$, then both denominators in \eqref{e:IncDiffCaseI} equal $\big(\Lambda+\beta_k\big)\big(\Lambda+\beta_k+\beta_l\big)$ and the increasing difference \eqref{e:IncDiffCaseI} is zero; (ii) if $B_k\leq x_k\leq x_k'$, then both denominators in \eqref{e:IncDiffCaseI} equal $\Lambda\big(\Lambda+\beta_l\big)$ and the increasing difference \eqref{e:IncDiffCaseI} is zero; If $x_k<B_k\leq x_k'$, then the increasing difference in \eqref{e:IncDiffCaseI} writes equivalently
\begin{align}\label{e:IncDiffCaseIbis}
\displaystyle\frac{{\mathds{1}_{\{x_i\geq B_i\}}}\beta_l}{\mathrm{denom}_1\mathrm{denom}_2}\big(\mathrm{denom}_2-\mathrm{denom}_1\big),
\end{align}
with $\mathrm{denom}_2=(\Lambda+\beta_k+\beta_l)(\Lambda+\beta_k)\geq \Lambda(\Lambda+\beta_l)=\mathrm{denom}_1$ as $\beta_j\geq 0$ for all $j\in 1 .. n$. Hence the increasing difference \eqref{e:IncDiffCaseI} is nonnegative.\\
\textit{Case $k=i$ and $l\neq i$:} Let $\mathbb{R}^2\ni(x_i,x_l)\overset{g}{\longmapsto} \frac{\mathds{1}_{\{x_i\geq B_i\}}}{\Lambda+\beta_i\mathds{1}_{\{x_i<B_i\}}+\beta_l\mathds{1}_{\{x_l<B_l\}}}\in\mathbb{R}_+$, with $\Lambda=\sum_{j\neq  i,l}\beta_j\mathds{1}_{\{x_j<B_j\}}$. For $x_i'\geq x_i$, $x_l'\geq x_l$, we form the increasing difference
\begin{align}\label{e:IncDiffCaseII}
\begin{split}
& g(x_i',x_l')-g(x_i',x_l)-g(x_i,x_l')+g(x_i,x_l) \\
&  \quad\quad= \displaystyle\frac{\mathds{1}_{\{x_i'\geq B_i\}}\beta_l\big(\mathds{1}_{\{x_l<B_l\}}-\mathds{1}_{\{x_l'<B_l\}}\big)}{\mathrm{denom}_1}-\displaystyle\frac{\mathds{1}_{\{x_i\geq B_i\}}\beta_l\big(\mathds{1}_{\{x_l<B_l\}}-\mathds{1}_{\{x_l'<B_l\}}\big)}{\mathrm{denom}_2},
\end{split}
\end{align}
with $\mathrm{denom}_1=\big(\Lambda+\beta_i\mathds{1}_{\{x_i'<B_i\}}+\beta_l\mathds{1}_{\{x_l'<B_l\}}\big)\big(\Lambda+\beta_i\mathds{1}_{\{x_i'<B_i\}}+\beta_l\mathds{1}_{\{x_l<B_l\}}\big)$ and $\mathrm{denom}_2=\big(\Lambda+\beta_i\mathds{1}_{\{x_i<B_i\}}+\beta_l\mathds{1}_{\{x_l'<B_l\}}\big)\big(\Lambda+\beta_i\mathds{1}_{\{x_i<B_i\}}+\beta_l\mathds{1}_{\{x_l<B_l\}}\big)$. If $x_l\leq x_l'<B_l$ or $B_l\leq x_l\leq x_l'$, then the increasing difference \eqref{e:IncDiffCaseII} is zero as the numerators of both terms are zero. If $x_l<B_l\leq x_l'$, then the numerator of the first terms in \eqref{e:IncDiffCaseII} is $\mathds{1}_{\{x_i'\geq B_i\}}$ and $\mathds{1}_{\{x_i\geq B_i\}}$ for the second term. In this case: (i) if $x_i\leq x_i'<B_i$, then both numerators in \eqref{e:IncDiffCaseII} are zero and the increasing difference \eqref{e:IncDiffCaseII} is zero; (ii) if $B_i\leq x_i\leq x_i'$, then both numerators in \eqref{e:IncDiffCaseII} are equal to $\mathds{1}_{\{x_i\geq B_i\}}\beta_l$ and both denominators are equal $\Lambda\big(\Lambda+\beta_l\big)$ so the increasing difference \eqref{e:IncDiffCaseII} is zero; If $x_i<B_i\leq x_i'$, then the second term in \eqref{e:IncDiffCaseII} is zero and the first term is nonnegative. Hence the increasing difference \eqref{e:IncDiffCaseII} is nonnegative.~$\square$\\

Hence the general results of 
Section \ref{s:CCR_SM} apply to $f_i$ and $g_i$
as per \eqref{e:LossFormCleargUncond}.
In particular,  for such $f_i$ and $g_i$, Propositions \ref{p:eclccva} and \ref{p:eckva} ensure the nondecreasing property of both CECL and EC w.r.t.\ covariance coefficients of $(\mathcal{X}_0,\mathbf{X})$ and $(\mathcal{X}_0,\mathbf{X}')$ under the elliptical setup \eqref{e:ModelElliptic}.
\subsection{CCP Use Case Design}\label{s:NumericsUseCaseCCP}
\begin{figure}[htp]
\begin{center}
\includegraphics[width=0.8\textwidth]{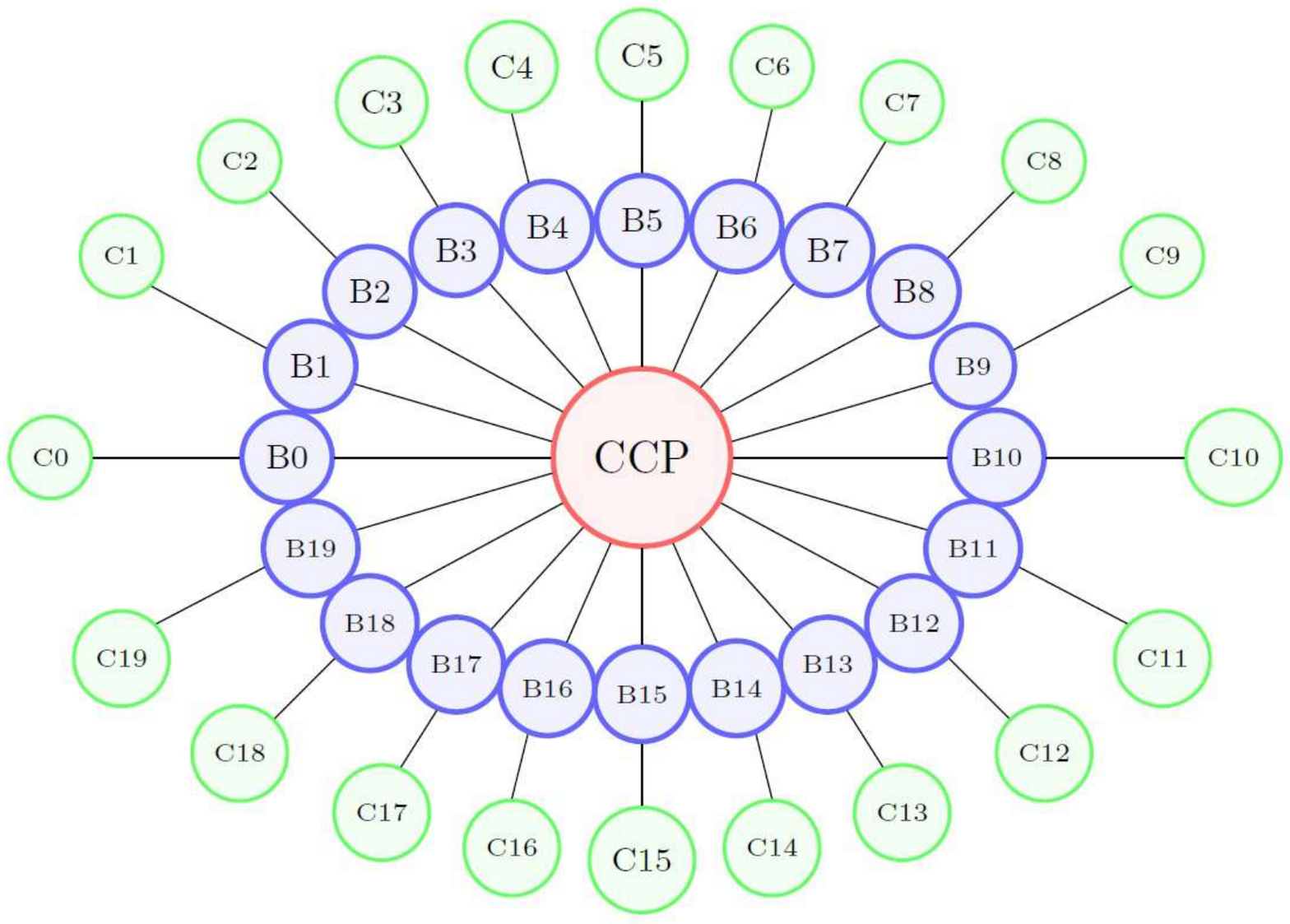}
\caption{Financial network composed of a CCP, its 20 members (labeled by B) and one cleared client per member.\label{fig:Network1CCP20Mbs}}
\end{center}
\end{figure}

We consider a CCP service with 20 members, labeled by $i\in 0.. n=19$, trading for cleared clients ({see \citet*[Sections 2 \& 5]{BastideCrepeyDrapeauTadese21a})}.  Each member faces one client. The corresponding financial network is depicted in Figure \ref{fig:Network1CCP20Mbs}.

All clients are assumed to be risk-free.
For any member $i$, its posted initial margin (IM) to the CCP is calculated based on the idea of a variation margin (VM) call not fulfilled over a slippage time period $\Delta_s$ at a confidence level $\alpha\in(1/2,1)$. {More precisely, the} IM {relies on} a VaR metric under the risk-neutral measure $\mathbb{Q}$ at the confidence level $\alpha$, denoted by $\mathbb{V}\mathrm{a}\mathbb{R}_\alpha$ applied to the non-coverage of VM call on the cleared portfolio. The latter {is modelled as} a scaled Student t-distribution $\mathcal{S}_{\nu}$ with $\nu$ degrees of freedom, with c.d.f.\ $S_{\nu}${. The} scaling reflects both $\Delta_s$, the portfolio nominal size ${\rm nom}_i$ and its standard deviation $\sigma_i$.
Namely, 
\begin{align}\label{e:IMCaldDef}
{\rm IM}_i=\mathbb{V}\mathrm{a}\mathbb{R}_\alpha \big({\rm nom}_i\sigma_i\sqrt{\Delta_s}\mathcal{S}_{\nu}\big)=|{\rm nom}_i|\,\sigma_i\sqrt{\Delta_s}{S_{\nu}}^{-1}(\alpha).
\end{align}
{A (funded)} default fund {pooled between its clearing members (see \citet*[Section 5]{BastideCrepeyDrapeauTadese21a})} is calculated at the CCP level as
\begin{align}\label{e:DF_size_Cover2}
{\rm Cover2} = {\rm SLOIM}_{(0)} + {\rm SLOIM}_{(1)},
\end{align}
for the two largest stressed losses over IM (${\rm SLOIM}_i$) among members, identified with subscripts $(0)$ and $(1)$. ${\rm SLOIM}_i$ is calculated as a value-at-risk $\mathbb{V}\mathrm{a}\mathbb{R}_{\alpha'}$ at a confidence level $\alpha'>\alpha$ of the loss over IM, i.e.
\begin{align}\label{e:SLOIMCaldDef}
{\rm SLOIM}_i=\mathbb{V}\mathrm{a}\mathbb{R}_{\alpha'}\big({\rm nom}_i\sigma_i\sqrt{\Delta_s}\mathcal{S}_{\nu}-{\rm IM}_i\big)= |{\rm nom}_i|\,\sigma_i\sqrt{\Delta_s}\Big({S_{\nu}}^{-1}(\alpha')-{S_{\nu}}^{-1}(\alpha)\Big).
\end{align}
The total amount \eqref{e:DF_size_Cover2} is then allocated between the clearing members to define their (funded) default fund contributions as ${\rm DF}_i=\frac{{\rm SLOIM}_i}{\sum_{j}{\rm SLOIM}_j}\times{\rm Cover2}$. Finally, the loss function of the reference member $0$ with default fund contribution ${\rm DF}_0$ {is of the form \eqref{e:LossFormCleargUncond} with $\beta_j=\frac{{\rm DF}_j}{{\rm DF}_0}$ and $m_i={\rm IM}_i+{\rm DF}_i$, i.e.,
\begin{align}\label{e:LossFormMbCCP}
\ell(\mathbf{X}) 
=\displaystyle\sum_{i=1}^n
\displaystyle\frac{1}{1+\displaystyle\sum_{j=1}^n \frac{{\rm DF}_j}{{\rm DF}_0}\mathds{1}_{\{X_j<B_j\}}} \mathds{1}_{\{X_i\geq B_i\}}\times\big(Y_i-{\rm IM}_i-{\rm DF}_i\big)^+.
\end{align}
}
Let $\mathrm{sgn}(x)=1$ if $x>0$, $0$ if $x=0$, $-1$ otherwise. An elliptical model is specified under $\mathbb{Q}$ as
\begin{align}\label{e:intermFactors}
\begin{cases}
Y_i={\rm nom}_i\sigma_i\sqrt{\Delta_l}\sqrt{\mathcal{K}}\left(\sqrt{\rho^{mkt}} \mathcal{E}+\sqrt{\rho^{wwr}}\mathcal{W}_i+\sqrt{1-\rho^{mkt}-\rho^{wwr}}\mathcal{E}_i\right) \\
X_i=\sqrt{\mathcal{K}}\left(\sqrt{\rho^{cr}} \mathcal{T}+\mathrm{sgn}({\rm nom}_i)\sqrt{\rho^{wwr}}\mathcal{W}_i
 +
\sqrt{1-\rho^{cr}-\rho^{wwr}}\mathcal{T}_i\right)
\end{cases}
\end{align}
for any $i\in 0\, ..\, n$, where $\mathcal{T}$, $\mathcal{T}_i$, $\mathcal{E}$, $\mathcal{E}_i$ and $\mathcal{W}_i$ are i.i.d. random variables following normal distributions and $\mathcal{K}/\nu$ follows an inverse-chi-squared distribution degree of freedom $\nu$, independent from any other considered r.v.. Hence all $X_i$'s and $Y_i$'s follow Student's t-distribution with degree of freedom $\nu$. $\Delta_l$ is the period accounting for the time taken by the CCP to novate or liquidate its portfolios in case of defaults (practically, $\Delta_l>\Delta_s$ by a few business days). $T$ represents the final maturity of the {CCP} portfolios. $B_i=\mathcal{S}_\nu^{-1}\left(1-{\rm DP}_i(T)\right)$ where ${\rm DP}_i (T)$ is the default probability over the period $[0,T]$ defined from a constant default intensity $\lambda_i$ given for each member $i$ in Table \ref{tab:MbNework1CCP20MbSetup} so that ${\rm DP}_i(T)=1-e^{-\lambda_i T}$ {which} can be obtained from their 1-year $\mathbb{Q}$ default probability ${\rm DP}_i (1Y)$, inferred either from the agency ratings or the CDS quotes when available, as {$\lambda_i=-\ln{\left(1-{\rm DP}_i (1Y)\right)}$.} The model is well defined if and only if $0<\rho^{wwr}<\min\left(1-\rho^{cr},1-\rho^{mkt}\right)$. Note that $\mathbb{C}{\rm ov}(X_i,Y_i)={\rm nom}_i\sigma_i\sqrt{\Delta_l}\sqrt{\rho^{wwr}}\mathrm{sgn}({\rm nom}_i)\sqrt{\rho^{wwr}}=|{\rm nom}_i|\sigma_i\sqrt{\Delta_l}\rho^{wwr}\geq 0$, hence increasing  $\rho^{wwr}$ leads to an increase of $\mathbb{C}{\rm ov}(X_i,Y_i)$.

The participants and portfolios input parameters are detailed in Table \ref{tab:MbNework1CCP20MbSetup}, where cm is the identifier of the clearing member, $\lambda$ is the one year $\mathbb{Q}$ default intensity of the member expressed in basis points, size represents the overall portfolio size of the member detained within the CCP, and vol is the annual volatility used for the portfolio variations. The portfolios listed in Table \ref{tab:MbNework1CCP20MbSetup} relate to the members towards the CCP (which are mirroring the portfolios between the members and their clients). The sizes of the CCP portfolios of members sum up to 0, in line with the CCP clearing condition (without proprietary trades).

\begin{remark}
The random variables \eqref{e:intermFactors} follow Student t-distributions that are continuous. Therefore, 0 is the only possible atom of the nonnegative credit loss \eqref{e:LossFormMbCCP}.
Hence, by \citet*[Corollary 5.3]{AT2002}, Definition \ref{d:ESdef} is equivalent to $\mathbb{ES}_{\alpha}(\mathcal{X}) ={\mathbb{E}} \left(\mathcal{X} | \mathcal{X}\geq {\mathbb{V}{\rm ar}_\alpha}(\mathcal{X}) \right)$ \citep[Eqn. (3.7)]{AT2002} whenever ${\mathbb{V}{\rm ar}_\alpha}(\mathcal{X}) >0$, i.e. for $\alpha\in\left(\frac{1}{2},1\right)$ sufficiently close to $1$ so that ${\mathbb{V}{\rm ar}_\alpha}(\mathcal{X}) >0$. In our numerical illustration with $\alpha=99.75\%$, this is indeed the case.  
\end{remark}

\begin{center}
\begin{table}[htp]
\begin{center}
\begin{minipage}{160mm}
\caption{Member characteristics and CCP portfolio parameters, ordered by decreasing \\ member $|$size$|$.}
{
\begin{center}
\begin{tabular}{@{}c}

\begin{tabular}{@{}lcccccccccc}
\hline
  cm & 0 & 1 & 2 & 3 & 4 & 5 & 6 & 7 & 8 & 9 \\ \hline
    $\lambda$ (bps) & 50 & 60 & 70 & 80 & 90 & 200 & 190 & 180 & 170 & 160\\
    size & -242 & 184 & 139 & 105 & -80 & -61 & -46 & 35 & 26 & -20 \\
    vol (\%) & 20 & 21 & 22 & 23 & 24 & 25 & 26 & 27 & 28 & 29 \\
\hline
\end{tabular} \\ \\

\begin{tabular}{@{}lcccccccccc}
\hline
  cm & 10 & 11 & 12 & 13 & 14 & 15 & 16 & 17 & 18 & 19\\ \hline
    $\lambda$ (bps) & 150 & 140 & 130 & 120 & 110 & 100 & 90 & 80 & 70 & 60\\
    size & -15 & -11 & -9 & -6 & 5 & -4 & -3 & 2 & 2 & -1\\
    vol (\%) & 30 & 31 & 32 & 33 & 34 & 35 & 36 & 37 & 38 & 39\\
\hline
\end{tabular}

\end{tabular}
\end{center}
}
\label{tab:MbNework1CCP20MbSetup}
\end{minipage}
\end{center}
\end{table}
\end{center}
\subsection{Numerical Results}\label{ss:NumResultsCCP}

The parameters of the CECL and EC calculations are summarized in Table \ref{tab:XVAconfig1CCP20Mbs}. The confidence level at $97\%$ for SLOIM in DF calibration allows for a ratio of default fund over initial margin of about $10\%$ in our calculations, a ratio (of this level or less) often observed in practice. Note that the chosen period length of $T=5$ years covers the bulk (if not the final maturity) of most realistic CCP portfolios.

\begin{table}[htp]
\begin{center}
\begin{minipage}{160mm}
\caption{CECL and EC calculation configuration.}
{
\begin{tabular}{@{}lc}
\hline
One-period length $T$ & 5 years \\ 
Liquidation period at default $\Delta_l$ & 5 days \\
Portfolio variations correlation $\rho^{mkt}$ & {4\%}\\
Credit factors correlation $\rho^{cr}$  & {5\%-95\%} \\
Correlation between credit factors and portfolio variations ${\rho^{wwr}_i}$&  {5\%-95\%} \\
IM covering period (MPoR) $\Delta_s$ & 2 days\\
IM quantile level & 95\% \\
SLOIM calculation for DF Cover-2 & VaR 97\%\\
DF allocation rule & based on IM\\
 Quantile level used for clearing members EC calculation & 99.75\%\\
Number of Monte-Carlo simulation (for CECL and EC computations) & {1M} \\
Number of batches (for EC computations) & 100 \\
\hline
\end{tabular}
}
\label{tab:XVAconfig1CCP20Mbs}
\end{minipage}
\end{center}
\end{table}

\begin{figure}[htp]
\begin{center}
\includegraphics[width=1\textwidth]{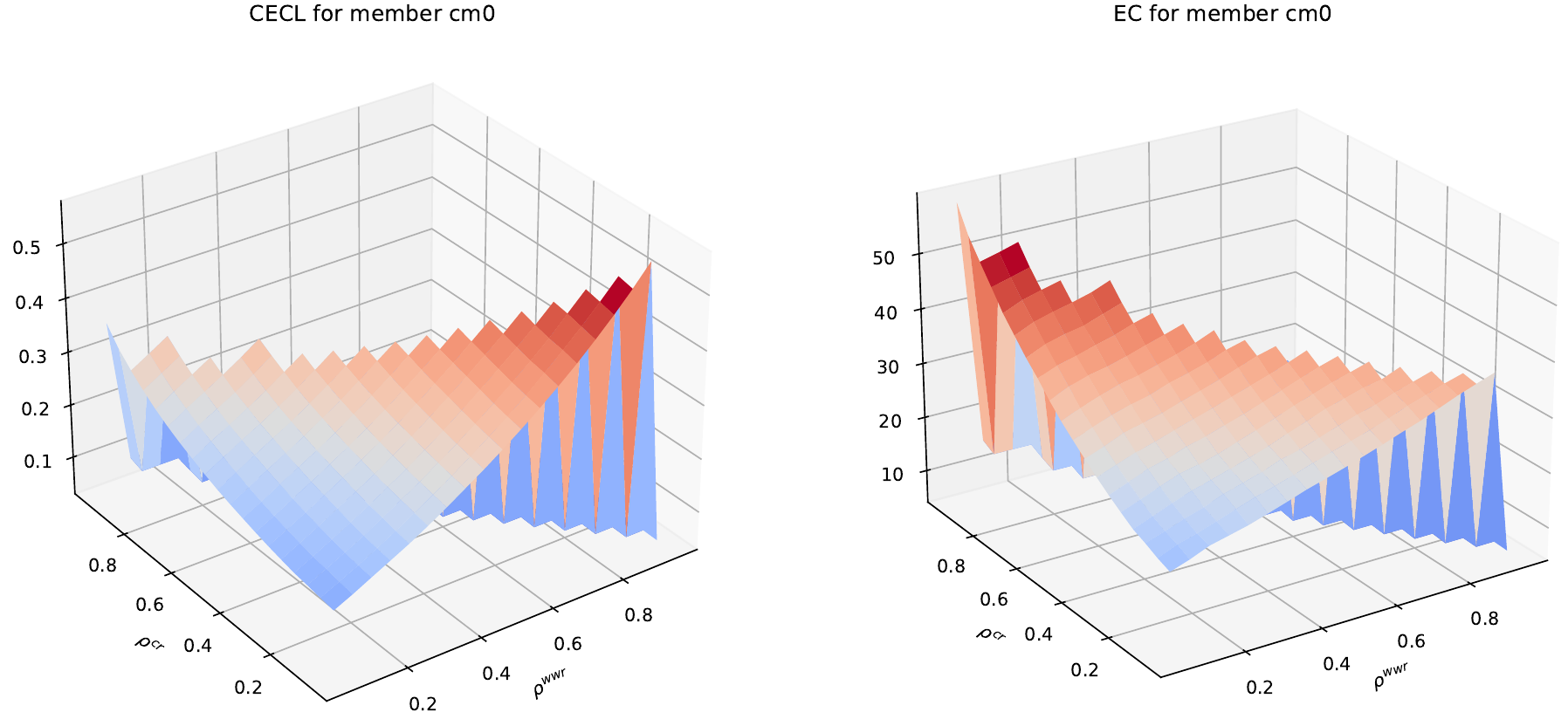}
\caption{Member 0 CECL and EC w.r.t.\ credit factors correlation $\rho^{cr}$ and credit and portfolio variation factors correlation $\rho^{wwr}$.\label{fig:CCVA_KVA_wrt_correl_cm0}}
\end{center}
\end{figure}

\begin{figure}[htp]
\begin{center}
\includegraphics[width=1\textwidth]{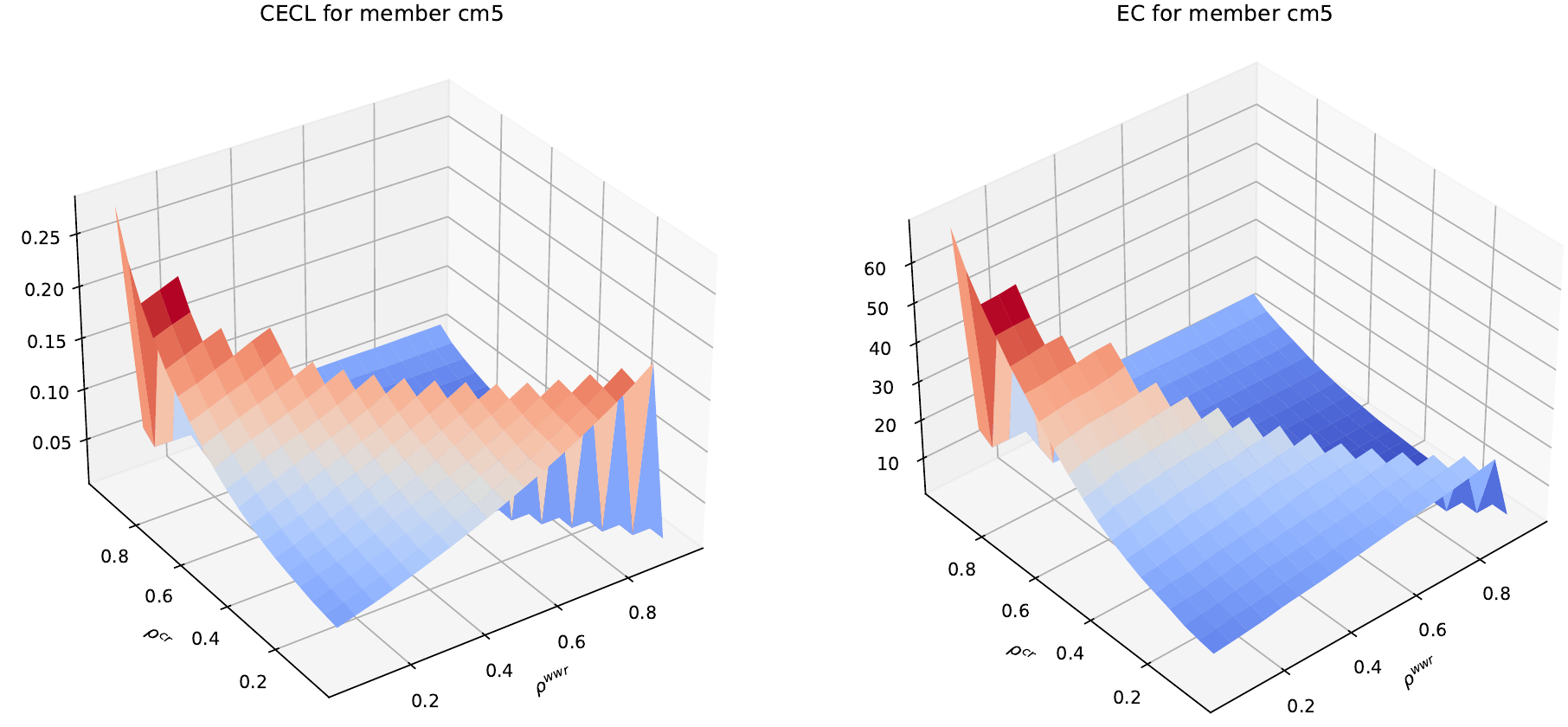}
\caption{Member 5 CECL and EC w.r.t.\ credit factors correlation $\rho^{cr}$ and credit and portfolio variation factors correlation  $\rho^{wwr}$.\label{fig:CCVA_KVA_wrt_correl_cm5}}
\end{center}
\end{figure}

\begin{figure}[htp]
\begin{center}
\includegraphics[width=1\textwidth]{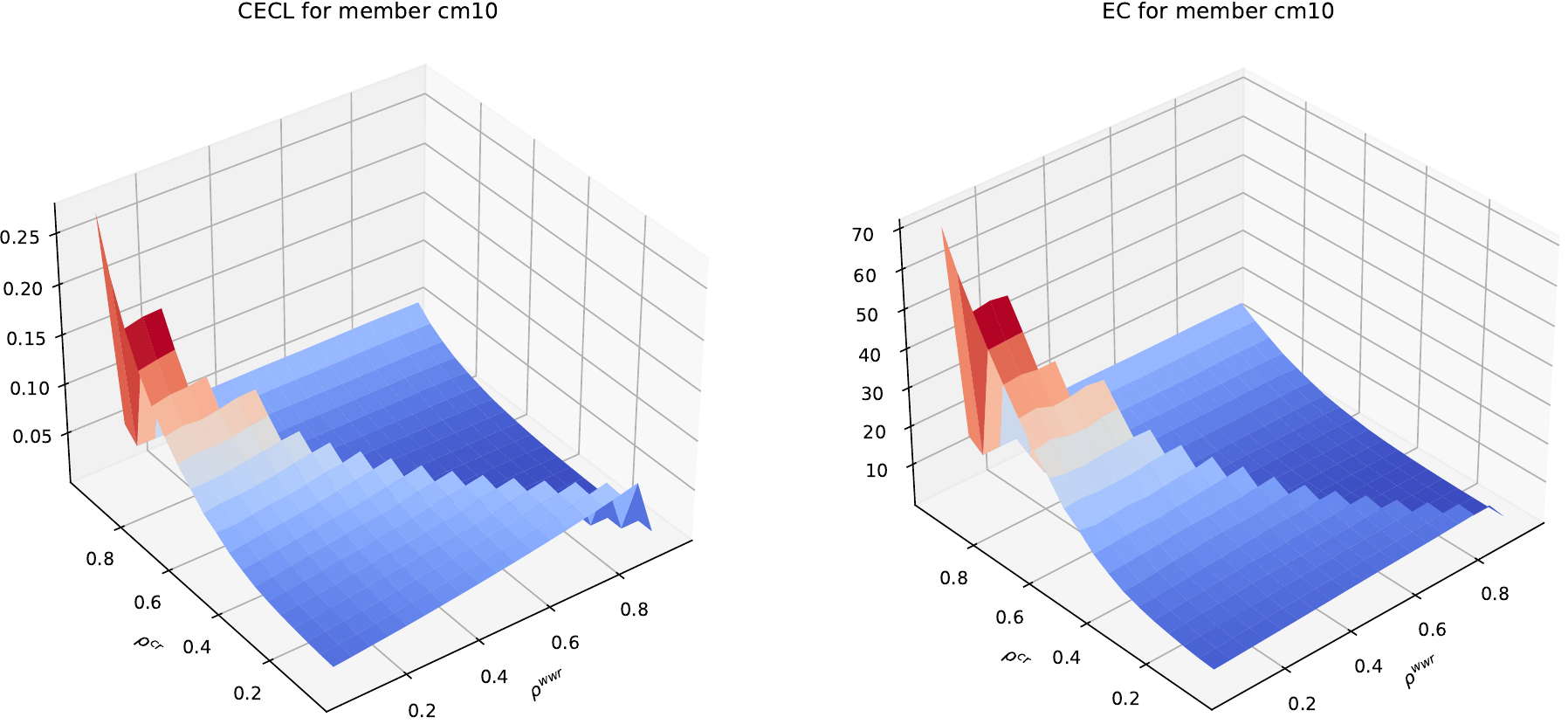}
\caption{Member 10 CECL and EC w.r.t.\ credit factors correlation $\rho^{cr}$ and credit and portfolio variation factors correlation $\rho^{wwr}$.\label{fig:CCVA_KVA_wrt_correl_cm10}}
\end{center}
\end{figure}

\begin{figure}[htp]
\begin{center}
\includegraphics[width=1\textwidth]{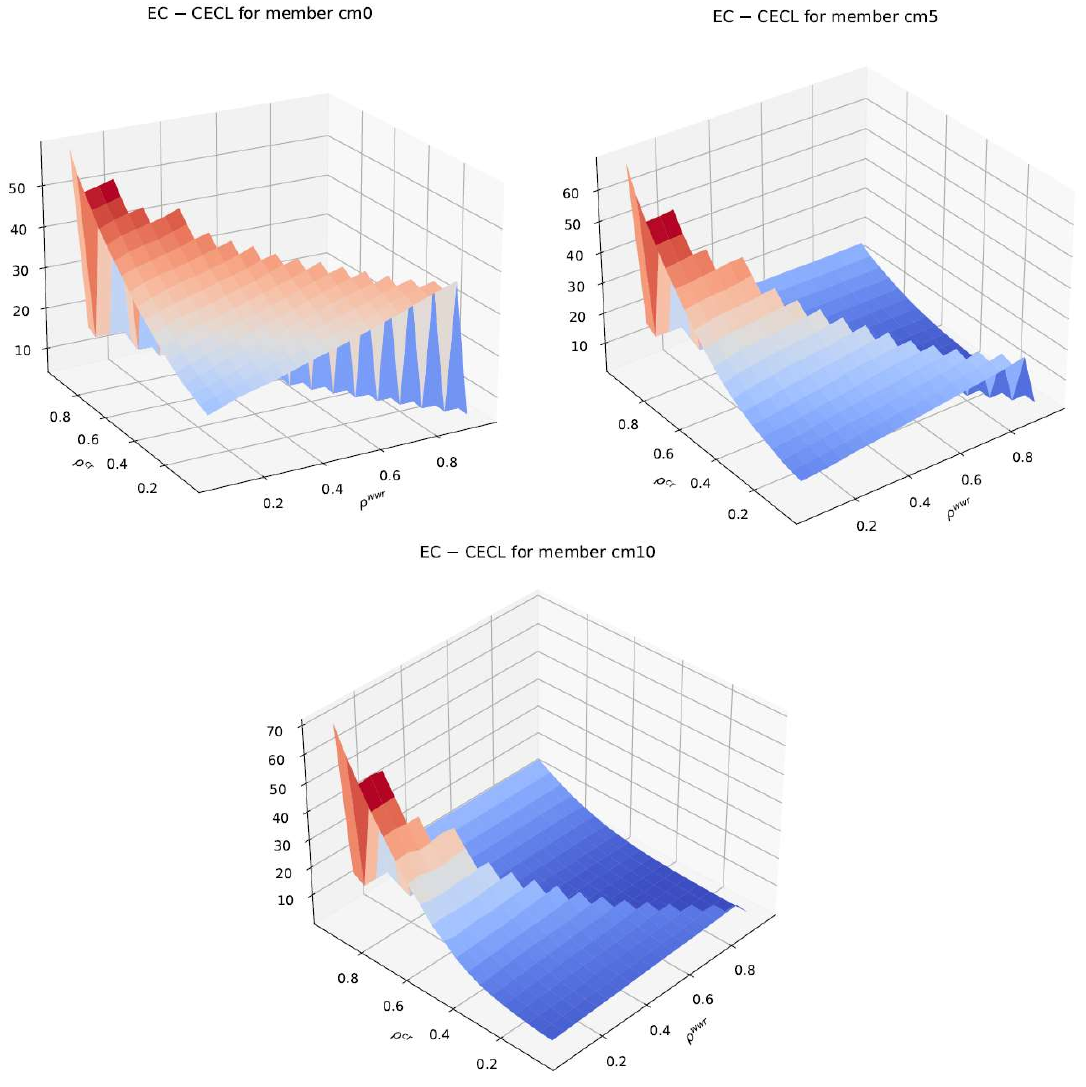}
\caption{Members 0, 5 and 10 ${\rm EC}-{\rm CECL}$ w.r.t.\ credit factors correlation $\rho^{cr}$ and credit and portfolio variation factors correlation $\rho^{wwr}$.\label{fig:CCVA_KVA_wrt_correl_cms_0_5_10}}
\end{center}
\end{figure}

Figures \ref{fig:CCVA_KVA_wrt_correl_cm0}, \ref{fig:CCVA_KVA_wrt_correl_cm5} and \ref{fig:CCVA_KVA_wrt_correl_cm10} show the results of CECL and EC calculated for the members 0, 5, and 10, each under their survival risk measure (i.e.\ letting them in turn play the role of the reference bank indexed by $0$ in { the above}). In each figure, the credit-credit correlation $\rho^{cr}$ and $\rho^{wwr}$ is  varied between 5\% and 95\%, using 5\% step. The same nondecreasing pattern is observed for all three members, with nonnegative incremental CECL and EC between two consecutive credit-credit and credit-market correlation steps, in line with Propositions \ref{p:eclccva} and \ref{p:eckva} applied under the specific case \eqref{e:intermFactors} of the elliptical setup \eqref{e:ModelElliptic}.
The market-market correlation $\rho^{mkt}$ has been kept constant with value 4\%. The results of the centered EC, i.e.\ ${\rm EC}-{\rm CECL}$, are also provided for each of these 3 members in Figure \ref{fig:CCVA_KVA_wrt_correl_cms_0_5_10}. As ${\rm CECL}\ll {\rm EC}$ holds for all three members, the monotonicity is also observed for this centered version of EC.

\begin{remark}
In our example, $\mathbb{C}{\rm ov}(Y_i,Y_j)={\rm nom}_i{\rm nom}_j\sigma_i\sigma_j\frac{\nu}{\nu-2}\rho^{mkt}$. Hence, depending on the sign of ${\rm nom}_i{\rm nom}_j$, increasing $\rho^{mkt}$ either increases or decreases $\mathbb{C}{\rm ov}(Y_i,Y_j)$. Thus, we cannot hope to observe a monotonous behaviour of ${\rm EC}$ or ${\rm CECL}$ w.r.t.\ $\rho^{mkt}$.
\end{remark}

\section{Credit Derivatives Supplement
}\label{s:CrDerivSM}
We analyse the monotonicity of default leg and coupon leg  of synthetic equity and senior CDO tranches prices w.r.t.\ credit correlation{, also using supermodular properties}. Such prices are obtained by taking the expected value under $\mathbb{Q}$, assumed to be, in this section, the risk-neutral pricing measure of the loss function underlying the CDO tranche contract. The characteristics of the payoff are as follows.
There are $n$ obligors, indexed by $i$. All underlying CDS are assumed to mature at some common time $T$. For any obligor $i$, the default time is denoted by $\tau_i$, the deterministic recovery rate is $R_i\in[0,1]$, the underlying notional is $N_i\geq 0$ and the loss given default is $L_i=(1-R_i)N_i$.
The maximum loss is $L_{\max}=\sum_{i=1}^n L_i$. To simplify calculations, we assume that the payments due to the obligors defaults are only made at maturity $T$ and the discounting rates are set to zero (nonzero discounting rates can be included as long as they are independent from the credit risk factors).

\begin{definition} The cumulative credit loss at time $t\leq T$ is  
\begin{align}\label{e:CumLoss}
L(t)=\sum_{i=1}^n L_i\mathds{1}_{\{\tau_i\leq t\}}.
\end{align}
The default leg of an equity tranche with maturity $T$ and detachment point $B\in(0,L_{\max}]$ is
\begin{align}\label{e:TranchedLossEQ}
\text{D}_{\text{eq}}(T) = L(T)-\big(L(T)-B\big)^+=\min\big(L(t),B\big).
\end{align}
The default leg of a senior tranche with maturity $T$ and attachment point $A\in[0,L_{\max})$ is
\begin{align}\label{e:TranchedLossSN}
\text{D}_{\text{sen}}(T) = \big(L(T)-A\big)^+.~\square
\end{align}
\end{definition}

\begin{definition}
The payment leg consists in payments, at $K$ regular times $t_k$ (with $t_K=T$), of a fixed spread $s$ applied to the remaining tranche amount at risk.
In the case of the equity tranche, the payment leg writes
\begin{align}\label{e:PmtLegEqTr}
\text{P}_{\text{eq}}(T)=s\displaystyle\frac{T}{K} \sum_{k=1}^K\big(B-L(t_k)\big)^+.
\end{align}
In the case of the senior tranche, it writes
\begin{align}\label{e:PmtLegSenTr}
\begin{array}{lcl}
\text{P}_{\text{sen}}(T) & = & s\displaystyle\frac{T}{K}\sum_{k=1}^K \left(L_{\max} - A - \big(L(t_k)-A\big)^+\right) \\
        & = & sT\big(L_{\max} - A\big)-s\displaystyle\frac{T}{K}\sum_{k=1}^K\big(L(t_k)-A\big)^+.~\square
\end{array}
\end{align}
\end{definition}

\noindent Putting default and payment payoffs together, we obtain, for the equity tranche,
\begin{align}\label{e:AllLegsEqTr}
\text{D}_{\text{eq}}(T)-\text{P}_{\text{eq}}(T)=L(T)-\big(L(T)-B\big)^+-s\displaystyle\frac{T}{K} \sum_{k=1}^K\big(B-L(t_k)\big)^+,
\end{align}
and, for the senior tranche,
\begin{align}\label{e:AllLegsSenTr}
\begin{array}{l} \text{D}_{\text{sen}}(T)-\text{P}_{\text{sen}}(T)=\\
\hspace{2cm}\big(L(T)-A\big)^+ -sT\big(L_{\max}+ A\big)+s\displaystyle\frac{T}{K}\sum_{k=1}^K\big(L(t_k)-A\big)^+.
\end{array}
\end{align}
Specifying $X_i=F^{-1}_i\big(1-\gamma_i(\tau_i)\big)$, where $F_i$ is the invertible c.d.f.\ of $X_i$, $\gamma_i$ the $\mathbb{Q}$ c.d.f.\ of $\tau_i$ and letting $B_i(t):=F^{-1}_i\big(1-\gamma_i(t)\big)$, we have $\{\tau_i\leq t\}=\{X_i\geq B_i(t)\}$ and
\begin{align}\label{e:CumLossExpr2}
L(t)=\displaystyle\sum_{i=1}^n L_i\mathds{1}_{\{X_i\geq B_i(t)\}}.
\end{align}
The following result precises the outlined application for comparing CDO tranche premiums in \citet*{CL08} with heterogeneous obligors under our static setup.
\begin{proposition}\label{pro:AppliArgtForCDO}
If $(X_1,\dots,X_n)\sim E_n({\boldsymbol \mu},\boldsymbol\Gamma,\psi)$ and $(X'_1,\dots,X'_n)\sim E_n({\boldsymbol \mu},\boldsymbol\Gamma',\psi)$, with $\boldsymbol\Gamma\leq\boldsymbol\Gamma'$ elementwise except for equal diagonal entries, then
\begin{align}\label{e:DLseniorLeg}
\mathbb{E}\left[\left(\displaystyle\sum_{i=1}^n L_i\mathds{1}_{\big\{X_i\geq B_i(t)\big\}}-A\right)^+\right]
\leq
\mathbb{E}\left[\left(\displaystyle\sum_{i=1}^n L_i\mathds{1}_{\big\{X'_i\geq B_i(t)\big\}}-A\right)^+\right],\,\,  A\in\mathbb{R}.
\end{align}
That is, the price of the default leg of a senior CDO tranche is nondecreasing w.r.t.\ the homogeneous correlation coefficient between the $X_i$'s noted $\rho^{cr}$; we also get that the price of the default leg of an equity CDO tranche is nonincreasing w.r.t.\ $\rho^{cr}$.
\end{proposition}
\proof
For any $t\in\mathbb{R}_+$, the function
\begin{align}\label{e:DetermLossLttVbFn}
\begin{array}{lccl}
     f_t: & \mathbb{R}^n & \longrightarrow & \mathbb{R}_+ \\
     & (x_1,\dots,x_n) & \longmapsto & \displaystyle\sum_{i=1}^n L_i\mathds{1}_{\{x_i\geq B_i(t)\}}
\end{array}
\end{align}
is nondecreasing  { on $\mathbb{R}^n$} and it is supermodular, by Lemma \ref{lem:aggLossRkSM}. Hence, due to the nondecreasing and convexity properties of $x\mapsto (x-A)^+$, \citet*[Theorem 3.9.3 f), page 113]{MS2002}, recalled in Section \ref{apdx:SMFns}, implies that $ (x_1,\dots,x_n)\mapsto\big(f_t(x_1,\dots,x_n)-A\big)^+$ is also nondecreasing supermodular. Moreover, by \citet*[Corollary 2.3]{BS88}, recalled in Section \ref{s:ellipt}, $(X_1,\dots,X_n)\leq_{sm}(X'_1,\dots,X'_n)$. Applying \citet[Definition 2.6]{M97} to $(X_1,\dots,X_n)$, $(X'_1,\dots,X'_n)$ and $ (x_1,\dots,x_n)\mapsto\big(f_t(x_1,\dots,x_n)-A\big)^+$ then yields 
the result for the senior tranche.\\ As $\mathbb{E}\left[\sum_{i=1}^n L_i\mathds{1}_{\big\{X_i\geq B_i(t)\big\}}\right]-\mathbb{E}\left[\left(\sum_{i=1}^n L_i\mathds{1}_{\big\{X_i\geq B_i(t)\big\}}-B\right)^+\right]$ is the price of an equity tranche default leg, where the left expectation term does not depend on $\rho^{cr}$, the result for the equity tranche follows.~$\square$

\begin{corollary}\label{cor:AppliArgtForCDO2}
Under the assumptions of Proposition \ref{pro:AppliArgtForCDO}, the price of the payment leg of the CDO equity (resp. senior) tranche is nondecreasing (resp. nonincreasing) w.r.t.\ $\rho^{cr}$.
\end{corollary}
\proof By call-put parity, 
\begin{align}\label{e:DLseniorLegCallPutPar}
\begin{array}{lcl}
     \mathbb{E}\left[\left(B-\displaystyle\sum_{i=1}^n L_i\mathds{1}_{\big\{X_i\geq B_i(t)\big\}}\right)^+\right] & = & \mathbb{E}\left[\left(\displaystyle\sum_{i=1}^n L_i\mathds{1}_{\big\{X'_i\geq B_i(t)\big\}}-B\right)^+\right]\\
     & & - B +\mathbb{E}\left[\displaystyle\sum_{i=1}^n L_i\mathds{1}_{\big\{X'_i\geq B_i(t)\big\}}\right]
\end{array}
\end{align}
so that, in view of \eqref{e:PmtLegEqTr}-\eqref{e:PmtLegSenTr}, as a consequence of Proposition \ref{pro:AppliArgtForCDO},
\begin{align}\label{e:DLseniorLegOpp}
\mathbb{E}\left[\left(B-\displaystyle\sum_{i=1}^n L_i\mathds{1}_{\big\{X_i\geq B_i(t)\big\}}\right)^+\right]
\leq
\mathbb{E}\left[\left(B-\displaystyle\sum_{i=1}^n L_i\mathds{1}_{\big\{X'_i\geq B_i(t)\big\}}\right)^+\right].~\square
\end{align}

\begin{remark} For mezzanine tranches, such results do not hold. Indeed, the tranched loss default leg payoff function
\begin{align}\label{e:TranchedLossMEZ}
    \text{D}_{\text{Mezz}}(T) = \big(L(T)-A\big)^+-\big(L(T)-B\big)^+,
\end{align}
where $A, B\in(0,L_{\max})$, is not a convex function of the cumulative loss, nor is the payment leg
\begin{align}\label{e:CpnLegMEZ}
    \begin{array}{lcl}
         \text{P}_{\text{Mezz}}(T) & = & s\displaystyle\frac{T}{K}\sum_{k=1}^K \Big[B - A - \Big(\big(L(t_k)-A\big)^+-\big(L(t_k)-B\big)^+\Big)\Big] \\
         & = & sT\big(B - A\big)-s\displaystyle\frac{T}{K}\sum_{k=1}^K\Big[\big(L(t_k)-A\big)^+-\big(L(t_k)-B\big)^+\Big].
    \end{array}
\end{align}
\end{remark}

\begin{figure}[htp]
\begin{center}
\includegraphics[width=1\textwidth]{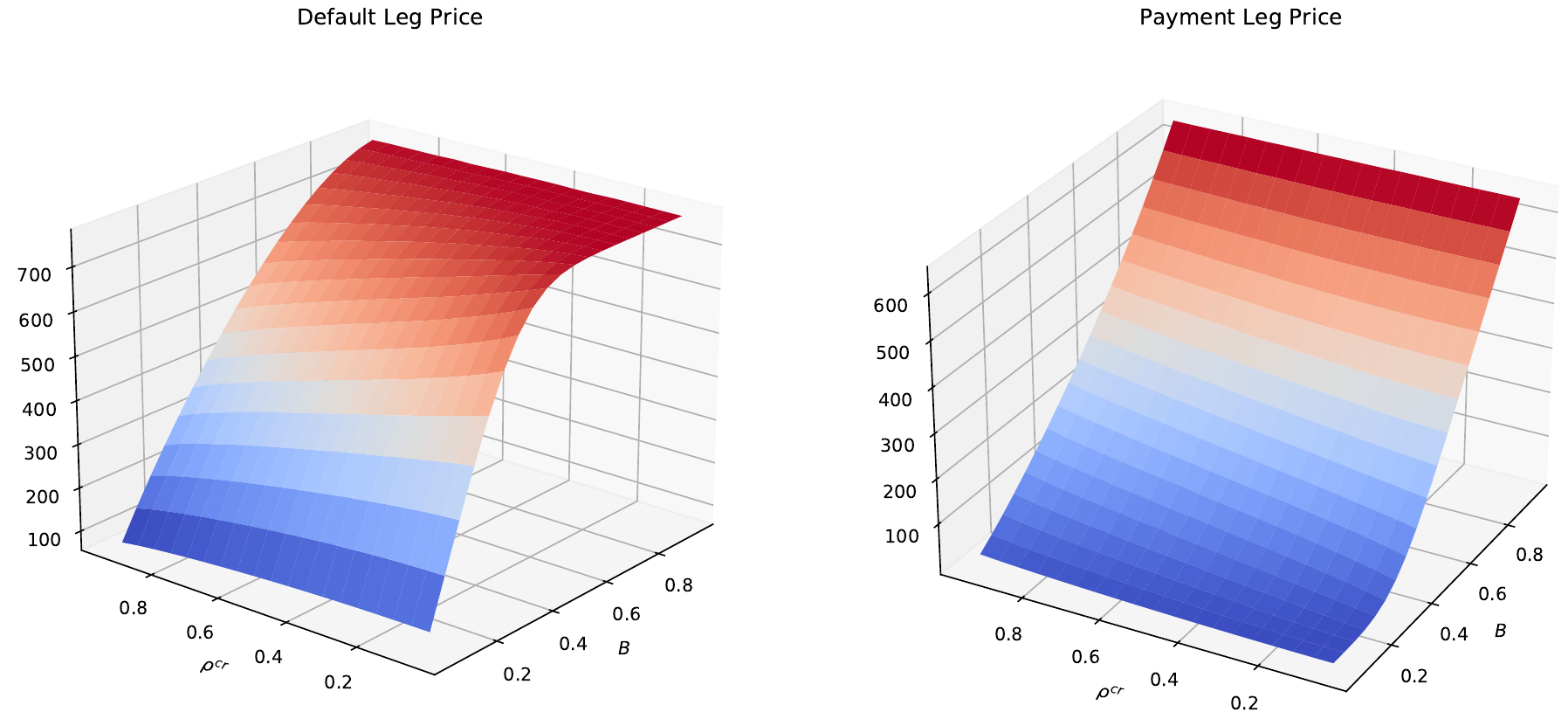}
\caption{Default leg and Payment leg prices of CDO equity tranches w.r.t.\ latent variable credit correlation $\rho^{cr}$ and detachment point $B$.\label{fig:EqTrancheDefaultLegRes}}
\end{center}
\end{figure}

\begin{figure}[htp]
\begin{center}
\includegraphics[width=1\textwidth]{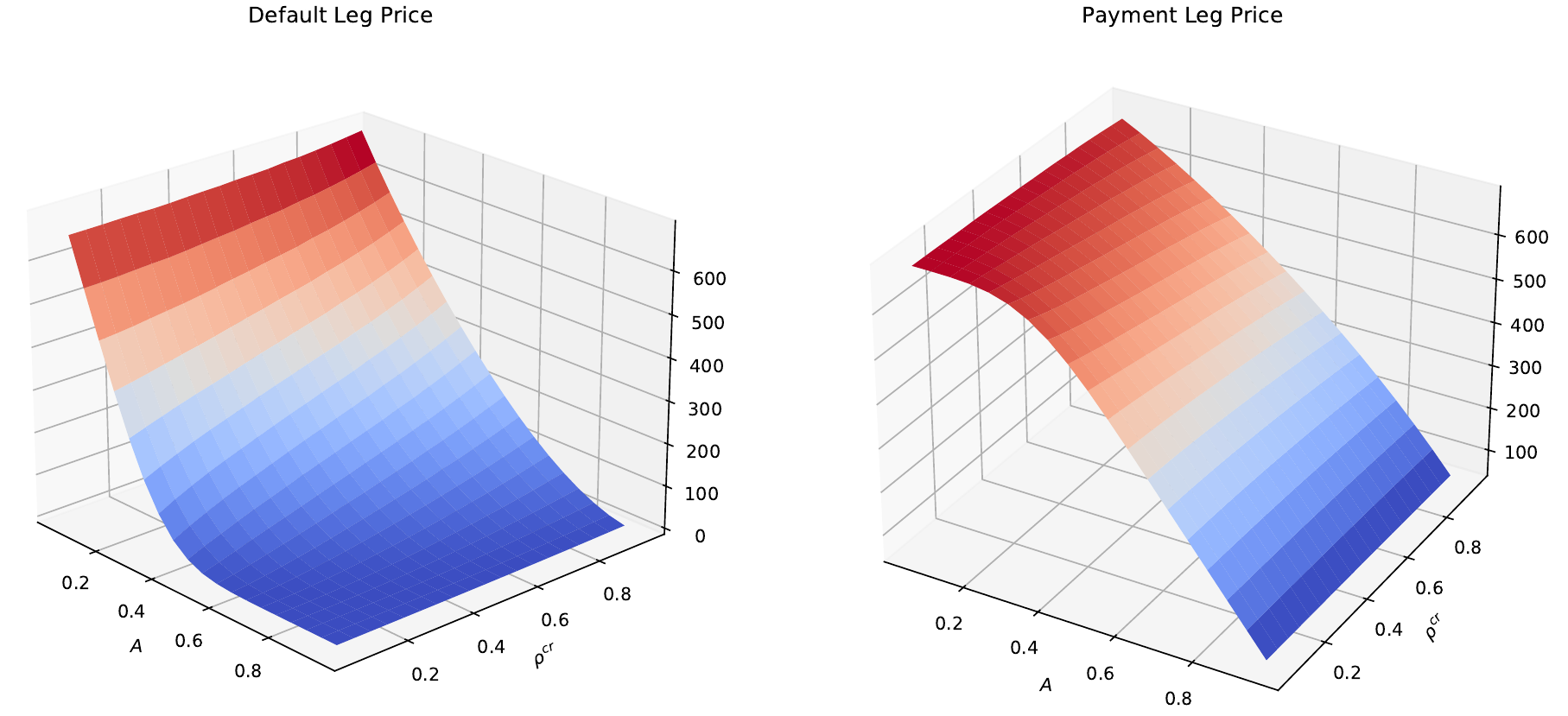}
\caption{Default leg and Payment leg prices of CDO senior tranches w.r.t.\ latent variable credit correlation $\rho^{cr}$ and attachment point $A$. Note that the axes are different from Figure \ref{fig:EqTrancheDefaultLegRes}, for a better readability.\label{fig:SenTrancheDefaultLegRes}}
\end{center}
\end{figure}

The results are illustrated in Figure \ref{fig:EqTrancheDefaultLegRes} for the equity tranches, varying detachment point from $5\%$ to $95\%$ with $5\%$ steps, i.e.\ considering the tranches from $[0,5\%]$ to $[0,95\%]$. Figure \ref{fig:SenTrancheDefaultLegRes} illustrates the results for the senior tranches, varying attachment point from $5\%$ to $95\%$ with $5\%$ steps, i.e.\ considering the tranches from $[5\%,100\%]$ to $[95\%,100\%]$. The correlation $\rho^{cr}$ is varied from $5\%$ to $95\%$ with $5\%$ step for both tranche types.
The parameters of the underlying obligors and CDSs are detailed in Table \ref{tab:CDOparams}, where values have been assigned arbitrarily to ensure heterogeneity of the various obligors. The CDO tranche spread has been set to $10\%$ with a single coupon paid at a maturity of $5$ years. The monotonicity patterns are observed for both tranches, with incremental prices between two consecutive credit correlation steps being nonpositive for the CDO equity tranche default leg prices and the CDO senior tranche payment leg prices. Incremental prices between two consecutive credit correlation steps are nonnegative for both the CDO equity tranche payment leg prices and the CDO senior tranche default leg prices. These results are in line with Proposition \ref{pro:AppliArgtForCDO} and Corollary \ref{cor:AppliArgtForCDO2}. Also, incremental prices between two attachment point steps are nonnegative for both the CDO equity tranche default leg and payment leg prices, as expected from \eqref{e:DLseniorLegOpp}. The incremental prices between two attachment point steps are nonpositive for the CDO senior tranche default leg and payment leg prices, as expected from \eqref{e:DLseniorLeg}.

\begin{table}[htp]
\begin{center}
\begin{minipage}{160mm}
\caption{CDO portfolios and obligors parameters.}
{
\begin{center}
\begin{tabular}{@{}c}

\begin{tabular}{@{}lcccccccccc}
\hline
  Obligor id & 1 & 2 & 3 & 4 & 5 & 6 & 7 & 8 & 9 & 10 \\
\hline
    Notional & 100 & 105 & 110 & 115 & 120 & 100 & 105 & 110 & 115 & 120 \\
    RR (\%) & 30 & 31 & 32 & 33 & 34 & 35 & 36 & 37 & 38 & 39 \\
    $\lambda (\%)$ & 2 & 2.5 & 3 & 3.5 & 4 & 4.5 & 5 & 5.5 & 6 & 6.5 \\
\hline
\end{tabular} \\ \\

\begin{tabular}{@{}lcccccccccc}
\hline
  Obligor id & 11 & 12 & 13 & 14 & 15 & 16 & 17 & 18 & 19 & 20 \\
\hline
    Notional & 100 & 105 & 110 & 115 & 120 & 100 & 105 & 110 & 115 & 120 \\ 
    RR (\%) & 40 & 30 & 31 & 32 & 33 & 34 & 35 & 36 & 37 & 38 \\
    $\lambda (\%)$ & 7 & 7.5 & 8 & 8.5 & 9 & 9.5 & 10 & 10.5 & 11 & 11.5 \\
\hline
\end{tabular} \\ \\

\begin{tabular}{@{}lcccccccccc}
\hline
  Obligor id & 21 & 22 & 23 & 24 & 25 & 26 & 27 & 28 & 29 & 30 \\
\hline
    Notional & 100 & 105 & 110 & 115 & 120 & 100 & 105 & 110 & 115 & 120 \\
    RR (\%) & 39 & 40 & 30 & 31 & 32 & 33 & 34 & 35 & 36 & 37 \\
    $\lambda (\%)$ & 12 & 12.5 & 13 & 13.5 & 14 & 14.5 & 15 & 15.5 & 16 & 16.5 \\
\hline
\end{tabular} \\

\end{tabular}
\end{center}
}
\label{tab:CDOparams}
\end{minipage}
\end{center}
\end{table}

\begin{remark}
    These results can be extended to stochastic recovery rates, correlated to default latent variables to capture wrong-way risk, in the spirit of \cite{FreyMcNeilNyfeler2001, FreyMcNeil2003, Pykhtin2003} and \cite{ChabaaneLaurentSalomon2004}.
\end{remark}

\section{Synthesis of the Results}\label{s:Conclusion}
The main mathematical results of the paper are summed up in Table \ref{tab:TheoreticalResults}. In a nutshell, if a participant uses a convex risk measure to assess its credit risk depicted as an aggregation of nonnegative losses driven by elliptically distributed factors, then the measure increases with the covariance coefficients between these factors. These results and their numerical illustrations support the use of such elliptical factor models for both risk management and regulatory credit provision and capital requirement purposes.

\begin{table}[htp]
\begin{center}
\begin{minipage}{160mm}
\caption{Main theoretical and applied results of the paper (with risk neutral measure $\mathbb{Q}$, reference participant labelled by $0$, related survival measure $\mathbb{Q}^0$).}
{
\begin{tabular}{ p{2.4cm}p{11.2cm} }
\hline
Proposition \ref{p:GenCaseSM}
& Let $f_i:\mathbb{R}^n\rightarrow\mathbb{R}$ be nondecreasing supermodular functions, $g_i:\mathbb{R}\rightarrow\mathbb{R}$ be nonnegative nondecreasing functions, $i\in 1\, ..\, n$. Then the function $(x_1,\dots,x_n,y_1,\dots,y_n) \overset{\ell}{\mapsto}
\sum_{i=1}^n f_i(x_1,\dots,x_n)g_i(y_i)$
is supermodular on $\mathbb{R}^{2n}$. \\ 
& \\
Theorem \ref{them:MonToSM}
& 
If $\mathbf{X}\leq_{sm^0}\mathbf{X}'$ and $\rho$ is a risk measure$^{\rm a}$ on $\mathfrak{X}^0$, then $\rho\left(\ell(\mathbf{X})\right)\leq \rho\left(\ell(\mathbf{X}')\right)$ holds whenever  $\ell(\mathbf{X}),\ell(\mathbf{X}')\in\mathfrak{X}^0$.
\\ \hline\hline
 Proposition \ref{pro:AppliArgtForCDO} and Corollary \ref{cor:AppliArgtForCDO2}
& The price of the default leg of an equity (resp. senior) tranche is nonincreasing (resp. nondecreasing) w.r.t.\ the credit correlation $\rho^{cr}$. The price of the payment leg of the CDO equity (resp. senior) tranche is nonincreasing (resp. nondecreasing) w.r.t.\ $\rho^{cr}$.\\
& \\
Proposition \ref{p:eclccva}
& Under Assumption \ref{a:HypCondSM}, ${\rm CECL}(\mathcal{X}_0,\mathbf{X})\leq {\rm CECL}(\mathcal{X}_0,\mathbf{X}')$, in particular in the elliptical setup \eqref{e:ModelElliptic}. \\ 
& \\
Proposition \ref{p:eckva}
& Under Assumption \ref{a:HypCondSM}, ${\rm EC}(\mathcal{X}_0,\mathbf{X})\leq {\rm EC}(\mathcal{X}_0,\mathbf{X}')$, in particular in the elliptical setup \eqref{e:ModelElliptic}. \\ 
\hline
\end{tabular}
}
\footnote{see Definition \ref{d:ConvexRiskMeas}.}
\label{tab:TheoreticalResults}
\end{minipage}
\end{center}
\end{table}


\bibliography{main}

\begin{thebibliography}{}

\bibitem[\protect\citeauthoryear{Acerbi and Tasche}{Acerbi and
  Tasche}{2002}]{AT2002}
Acerbi, C. and D.~Tasche (2002).
\newblock On the coherence of expected shortfall.
\newblock {\em Journal of Banking \& Finance\/}~{\em 26}, 1487--1503.

\bibitem[\protect\citeauthoryear{Albanese, Cr\'epey, Hoskinson, and
  Saadeddine}{Albanese et~al.}{2021}]{CrepeyHoskinsonSaadeddine2019}
Albanese, C., S.~Cr\'epey, R.~Hoskinson, and B.~Saadeddine (2021).
\newblock {XVA} analysis from the balance sheet.
\newblock {\em Quantitative Finance\/}~{\em 21\/}(1), 99--123.

\bibitem[\protect\citeauthoryear{Bastide, Cr\'epey, Drapeau, and
  Tadese}{Bastide et~al.}{2023}]{BastideCrepeyDrapeauTadese21a}
Bastide, D., S.~Cr\'epey, S.~Drapeau, and M.~Tadese (2023).
\newblock Derivatives risks as costs in a one-period network model.
\newblock {\em Frontiers of Mathematical Finance\/}~{\em 2\/}(3), 283--312.

\bibitem[\protect\citeauthoryear{B\"auerle and M\"uller}{B\"auerle and
  M\"uller}{1998}]{BM98}
B\"auerle, N. and A.~M\"uller (1998).
\newblock Modelling and comparing dependencies in multivariate risk portfolios.
\newblock {\em ASTIN Bulletin\/}~{\em 28\/}(1), 59--76.

\bibitem[\protect\citeauthoryear{B\"auerle and M\"uller}{B\"auerle and
  M\"uller}{2006}]{BM05}
B\"auerle, N. and A.~M\"uller (2006).
\newblock Stochastic orders and risk measures: consistency and bounds.
\newblock {\em Insurance: Mathematics and Economics\/}~{\em 38\/}(1), 132--148.

\bibitem[\protect\citeauthoryear{Block and Sampson}{Block and
  Sampson}{1988}]{BS88}
Block, H.~W. and A.~R. Sampson (1988).
\newblock Conditionally ordered distributions.
\newblock {\em Journal of Multivariate Analysis\/}~{\em 27}, 91--104.

\bibitem[\protect\citeauthoryear{Brigo, Morini, and Pallavicini}{Brigo
  et~al.}{2013}]{Brigo2013}
Brigo, D., M.~Morini, and A.~Pallavicini (2013).
\newblock {\em Counterparty Credit Risk, Collateral and Funding: With Pricing
  Cases for All Asset Classes}.
\newblock Wiley.

\bibitem[\protect\citeauthoryear{Chabaane, Laurent, and Salomon}{Chabaane
  et~al.}{2004}]{ChabaaneLaurentSalomon2004}
Chabaane, A., J.-P. Laurent, and J.~Salomon (2004, 03).
\newblock Double impact: Credit risk assessment and collateral value.
\newblock {\em Revue Finance\/}~{\em 25}.

\bibitem[\protect\citeauthoryear{Cousin and Laurent}{Cousin and
  Laurent}{2008}]{CL08}
Cousin, A. and J.-P. Laurent (2008).
\newblock Comparison results for exchangeable credit risk portfolios.
\newblock {\em Insurance: Mathematics and Economics\/}~{\em 42\/}(3),
  1118--1127.

\bibitem[\protect\citeauthoryear{{European~Central~Bank}}{{European~Central~Bank}}{2019}]{ECB2019}
{European~Central~Bank} (October 2019).
\newblock {ECB} guide to internal models.
\newblock Available at
  \url{https://www.bankingsupervision.europa.eu/ecb/pub/pdf/ssm.guidetointernalmodels_consolidated_201910~97fd49fb08.en.pdf}.
  Accessed on October 14, 2021.

\bibitem[\protect\citeauthoryear{Fang, Kotz, and Ng}{Fang
  et~al.}{1990}]{FKN1990}
Fang, K.-T., S.~Kotz, and K.~W. Ng (1990).
\newblock {\em Symmetric Multivariate and Related Distributions}.
\newblock CRC Press, New York.

\bibitem[\protect\citeauthoryear{F\"{o}llmer and Schied}{F\"{o}llmer and
  Schied}{2016}]{follmerSchied2016}
F\"{o}llmer, H. and A.~Schied (2016).
\newblock {\em Stochastic Finance: An Introduction in Discrete Time\/} (4th
  ed.).
\newblock De Gruyter Graduate, Berlin, Germany.

\bibitem[\protect\citeauthoryear{Frey and McNeil}{Frey and
  McNeil}{2003}]{FreyMcNeil2003}
Frey, R. and A.~McNeil (2003, 07).
\newblock Dependent defaults in models of portfolio credit risk.
\newblock {\em Journal of Risk\/}~{\em 6}.

\bibitem[\protect\citeauthoryear{Frey, McNeil, and Nyfeler}{Frey
  et~al.}{2001}]{FreyMcNeilNyfeler2001}
Frey, R., A.~McNeil, and M.~Nyfeler (2001, 01).
\newblock Copulas and credit models.
\newblock {\em Risk\/}~{\em 14}, 111--114.

\bibitem[\protect\citeauthoryear{Gregory}{Gregory}{2015}]{gregory2015xva}
Gregory, J. (2015).
\newblock {\em The xVA challenge: counterparty credit risk, funding, collateral
  and capital}.
\newblock Wiley.

\bibitem[\protect\citeauthoryear{Kallenberg}{Kallenberg}{2021}]{Kallenberg2021}
Kallenberg, O. (2021).
\newblock {\em Foundations of Modern Probability\/} (3 ed.).
\newblock Switzerland: Springer Nature.

\bibitem[\protect\citeauthoryear{Marshall and Olkin}{Marshall and
  Olkin}{1979}]{MarshallOlkinArnold1979}
Marshall, A.~W. and I.~Olkin (1979).
\newblock {\em Inequalities: Theory of Majorization and Its Applications}.
\newblock Academic Press, New York.

\bibitem[\protect\citeauthoryear{McNeil, Frey, and Embrechts}{McNeil
  et~al.}{2015}]{McNeilFreyEmbrechts2015}
McNeil, A.~J., R.~Frey, and P.~Embrechts (2015).
\newblock {\em Quantitative Risk Management: Concepts, Techniques and Tools\/}
  (revised ed.).
\newblock Princeton Series in Finance.

\bibitem[\protect\citeauthoryear{Meyer and Strulovici}{Meyer and
  Strulovici}{2013}]{MS2013}
Meyer, M.~A. and B.~Strulovici (2013, May).
\newblock {The Supermodular Stochastic Ordering}.
\newblock CEPR Discussion Papers 9486.

\bibitem[\protect\citeauthoryear{Meyer and Strulovici}{Meyer and
  Strulovici}{2012}]{MS2012}
Meyer, M.~A. and B.~Strulovici (July 2012).
\newblock Increasing interdependence of multivariate distributions.
\newblock {\em Journal of Economic Theory\/}~{\em 147\/}(4), 91--104.

\bibitem[\protect\citeauthoryear{Meyer-Nieberg}{Meyer-Nieberg}{1991}]{MeyerNieberg1991}
Meyer-Nieberg, P. (1991).
\newblock {\em Banach Lattices}.
\newblock Universitext. Springer-Verlag.

\bibitem[\protect\citeauthoryear{M\"uller}{M\"uller}{1997}]{M97}
M\"uller, A. (1997).
\newblock Stop-loss order for portfolios of dependent risks.
\newblock {\em Insurance: Mathematics and Economics\/}~{\em 21}, 219--223.

\bibitem[\protect\citeauthoryear{M\"uller and Scarsini}{M\"uller and
  Scarsini}{2000}]{MS2000}
M\"uller, A. and M.~Scarsini (2000).
\newblock Some remarks on the supermodular order.
\newblock {\em Journal of Multivariate Analysis\/}~{\em 73}, 107--119.

\bibitem[\protect\citeauthoryear{M\"uller and Stoyan}{M\"uller and
  Stoyan}{2002}]{MS2002}
M\"uller, A. and D.~Stoyan (2002).
\newblock {\em Comparison Methods for Stochastic Models and Risks}.
\newblock Wiley Series in Probability and Statistics.

\bibitem[\protect\citeauthoryear{Pflug and R\"{o}misch}{Pflug and
  R\"{o}misch}{2007}]{PflugWerner2007}
Pflug, G.~C. and W.~R\"{o}misch (2007).
\newblock {\em Modeling, Measuring and Managing Risk}.
\newblock World Scientific Publishing Co. Pte. Ltd., Singapore.

\bibitem[\protect\citeauthoryear{Pykhtin}{Pykhtin}{2003}]{Pykhtin2003}
Pykhtin, M. (2003, 01).
\newblock Unexpected recovery risk.
\newblock {\em Risk\/}~{\em 16}, 74--78.

\bibitem[\protect\citeauthoryear{Ruszczy\'{n}ski and Shapiro}{Ruszczy\'{n}ski
  and Shapiro}{2006}]{RS2006}
Ruszczy\'{n}ski, A. and A.~Shapiro (2006).
\newblock Optimization of convex risk functions.
\newblock {\em Mathematics of Operations Research\/}~{\em 31\/}(3), 433--452.

\bibitem[\protect\citeauthoryear{Shaked and Shanthikumar}{Shaked and
  Shanthikumar}{1997}]{SS97}
Shaked, M. and J.~G. Shanthikumar (1997).
\newblock Supermodular stochastic orders and positive dependence of random
  vectors.
\newblock {\em Journal of Multivariate Analysis\/}~{\em 61\/}(MV971656),
  86--101.

\bibitem[\protect\citeauthoryear{Shapiro}{Shapiro}{2013}]{Shapiro2013}
Shapiro, A. (2013, February).
\newblock {On Kusuoka Representation of Law Invariant Risk Measures}.
\newblock {\em Mathematics of Operations Research\/}~{\em 38\/}(1), 142--152.

\bibitem[\protect\citeauthoryear{{Single~Resolution~Board}}{{Single~Resolution~Board}}{2021}]{SRB2021}
{Single~Resolution~Board} (2021, 01).
\newblock {The~Single~Resolution~Fund}.
\newblock Technical report, {Single~Resolution~Board}.
\newblock Available on
  \url{https://www.srb.europa.eu/en/content/single-resolution-fund}. Accessed
  on September 8, 2022.

\bibitem[\protect\citeauthoryear{Steerneman and van Perlo-ten Kleij}{Steerneman
  and van Perlo-ten Kleij}{2005}]{Steerneman2005}
Steerneman, A. and F.~van Perlo-ten Kleij (2005, September).
\newblock Spherical distributions: Schoenberg (1938) revisited.
\newblock {\em Expositiones Mathematicae\/}~{\em 23\/}(3), 281--287.

\bibitem[\protect\citeauthoryear{Yildiz}{Yildiz}{2010}]{Yildiz2010}
Yildiz, M. (2010, April).
\newblock Lecture notes on supermodular games.
\newblock Available at \url{https://dspace.mit.edu/handle/1721.1/106995/}.
  Accessed on December 22, 2021.

\end{thebibliography}

\bigskip

\appendix
\section{Supermodular Functions}\label{apdx:SMFns}
\noindent\textit{\citet[Definitions 2.4 and 2.6]{M97}}\label{def:SuperModFuncRn}
A function $f:\mathbb{R}^n\longrightarrow\mathbb{R}$ is said to be supermodular if 
\begin{align}\label{e:SupModFunc}
\begin{split}
& f(x_1,\dots,x_i+\varepsilon,\dots,x_j+\delta,\dots,x_n)-f(x_1,\dots,x_i+\varepsilon,\dots,x_j,\dots,x_n)\geq \\ 
& \quad\quad\quad f(x_1,\dots,x_i,\dots,x_j+\delta,\dots,x_n)-f(x_1,\dots,x_i,\dots,x_j,\dots,x_n)
\end{split}
\end{align}
holds for all $x\in\mathbb{R}^n$, $\varepsilon, \delta >0$ and $1\leq i<j\leq n$.~$\square$\\

\noindent\textit{\citet[Theorem 2.2 (a)]{MS2000}}
For functions twice differentiable on $\mathbb{R}^d$, the supermodularity is equivalent to the nonnegativity of its second derivatives.~$\square$\smallskip

\vspace{0.3cm}
\noindent More general definitions can be found in \citet*{MeyerNieberg1991} and \citet*{Yildiz2010}.\smallskip

\vspace{0.3cm}
\noindent\textit{\citet[Section 1.1, page 1]{MeyerNieberg1991}}\label{def:LatticeSet}
A partially ordered set $(\mathfrak{S},\geq)$ is said to be a lattice if and only if any two elements $x,y$ have a greatest common minorant, denoted $x\wedge y$, and a least common majorant, denoted $x\vee y$.~$\square$

\vspace{0.3cm}
\noindent\textit{\citet[Definition 6, page 6]{Yildiz2010}}\label{def:GenSuperModFunc}
A function $f: \mathfrak{S} \longrightarrow\mathbb{R}$ is said to be supermodular on a lattice $(\mathfrak{S},\geq)$ if
\begin{equation}\label{e:GenSuperModFunc}
f(x\vee y)+ f (x\wedge y) \geq f(x)+ f(y)
\end{equation}
holds for all $x,y \in \mathfrak{S}$.~$\square$

\vspace{0.3cm}
\noindent For a family of lattices $(\mathfrak{S}_1,\leq),\dots,(\mathfrak{S}_n,\leq)$, let $\mathfrak{S}=\mathfrak{S}_1\times\dots\times\mathfrak{S}_n$ be endowed with the coordinate-wise order $(x_1,\dots,x_n)\leq (y_1,\dots,y_n)$ if and only if $\forall i, x_i \leq y_i$. This order makes $\mathfrak{S}=\mathfrak{S}_1\times\dots\times\mathfrak{S}_n$ a lattice. For $x\in\mathfrak{S}$ and any $i$ and $j$, define $\mathbf{x}_{-(i,j)}=(x_k)_{k\neq i,j}$. For any function $f:\mathfrak{S}\longrightarrow \mathbb{R}$, define $f(\cdot|\mathbf{x}_{-(i,j)}):\mathfrak{S}_i\times\mathfrak{S}_j \longrightarrow\mathbb{R}$ as the restriction of $f$ to vectors with entries other than $i$ and $j$ fixed at $\mathbf{x}_{-(i,j)}$ \citep[Section 2.4]{Yildiz2010}.\smallskip\smallskip

\noindent\textit{\citet[remark after Definition 7, page 7]{Yildiz2010}}
A function $f:\mathfrak{S}=\mathfrak{S}_1\times\dots\times\mathfrak{S}_n\longrightarrow\mathbb{R}$ is said to be pairwise supermodular if
\begin{equation}\label{e:PairSuperModFunc}
f\big((x_i,x_j)\vee(x'_i,x'_j)|\mathbf{x}_{-(i,j)}\big)+f\big((x_i,x_j)\wedge(x'_i,x'_j)|\mathbf{x}_{-(i,j)}\big)\geq f\big(x_i,x_j|\mathbf{x}_{-(i,j)}\big)+f\big(x'_i,x'_j|\mathbf{x}_{-(i,j)}\big)
\end{equation}
holds for all $x_1,\dots,x_n\in\mathfrak{S}_1\times\dots\mathfrak{S}_n$.~$\square$

\vspace{0.3cm}
\noindent\textit{\citet[Definition 7]{Yildiz2010}}\label{def:IncDiffFunc}
A function $f:\mathfrak{S}=\mathfrak{S}_1\times\dots\times\mathfrak{S}_n\longrightarrow\mathbb{R}$ is said to have increasing differences if
\begin{align}\label{e:IncrDiffDef}
&\big[x_i\geq x_{i'},\, x_j\geq x_{j'}\big]\Longrightarrow\\&\quad\quad\quad f(x_{i'},x_{j'}|\mathbf{x}_{-(i,j)})-f(x_{i'},x_j|\mathbf{x}_{-(i,j)})\geq f(x_i,x_{j'}|\mathbf{x}_{-(i,j)})-f(x_i,x_j|\mathbf{x}_{-(i,j)})
\end{align}
holds for any $\mathbf{x}=(x_1,\dots,x_n)$, $i$, $j$, $i'$, $j'$.~$\square$

\vspace{0.3cm}
\noindent If the partial order is a total order, increasing differences and supermodularity are equivalent. In particular:\smallskip

\noindent\textit{\citet[Corollary 1]{Yildiz2010}}\label{cor:SMequivDefs}
For any $f:\mathbb{R}^n\longrightarrow \mathbb{R}$, the following are equivalent:
\begin{enumerate}
    \item f is supermodular;
    \item f has increasing differences;
    \item f is pairwise supermodular.~$\square$
\end{enumerate}

\noindent\textit{\citet[Theorem 3.9.3 f), page 113]{MS2002}}\label{th:composCvxIncSM}
If $f:\mathbb{R}^n\longrightarrow \mathbb{R}$ is nondecreasing and supermodular and $\phi: \mathbb{R}\longrightarrow\mathbb{R}$ is nondecreasing and convex, then $\phi\circ f:\mathbb{R}^n\longrightarrow \mathbb{R}$ is nondecreasing supermodular.~$\square$

The following two Lemmas \ref{lem:aggLossRkSM} and \ref{l:crossVbSM} will be the building blocks for establishing our main result Theorem \ref{them:MonToSM}.

\begin{lemma}\label{lem:aggLossRkSM}
The supermodularity property is satisfied by any function $\mathbb{R}^n\ni(x_1,\dots,x_n)\mapsto h(x_1,\dots,x_n)\in\mathbb{R}$ that can be written as $h(x_1,\dots,x_n)=\sum_{i=1}^n h_i(x_i)$ for some functions $h_1,\dots,h_n$ of single arguments. In particular, for any constants $B_1,\dots,B_n$, the function $(x_1,\dots,x_n)\longmapsto{ \sum_{i=1}^n\mathds{1}_{\{x_i\geq B_i\}}}$ is supermodular. This function is also nondecreasing { on $\mathbb{R}^n$}.
\end{lemma}
\proof
By \citet[Corollary 1]{Yildiz2010}, recalled in Section \ref{apdx:SMFns}, we can focus on increasing differences. Let $h(x_i, x_j|\mathbf{x}_{-i,j})$ denote the function $h$ applied to $x_i$ and $x_j$ but keeping all other arguments $\mathbf{x}_{-i,j}:=\left(x_k\right)_{k\neq i,j}$ fixed.
Fixing $\delta, \varepsilon>0$, the difference $h(x_i+\delta,x_j+\varepsilon|\mathbf{x}_{-i,j})-h(x_i,x_j+\varepsilon|\mathbf{x}_{-i,j})-h(x_i+\delta,x_j|\mathbf{x}_{-i,j})+h(x_i,x_j|\mathbf{x}_{-i,j})=h_i(x_i+\delta)+h_j(x_j+\varepsilon)-h_i(x_i)-h_j(x_j+\varepsilon)-h_i(x_i+\delta)-h_j(x_j)+h_i(x_i)+h_j(y_j)$ simplifies to $0$, showing the supermodularity of $h$.~$\square$
\begin{lemma}\label{l:crossVbSM}
If $g:\mathbb{R}\rightarrow\mathbb{R}$ and $h:\mathbb{R}\rightarrow\mathbb{R}$ are both nondecreasing, then $(x,y)\mapsto g(x)h(y)$ is supermodular.
\end{lemma}
\proof
If $x'\geq x$ and $y'\geq y$, then
\begin{align}
\begin{split}
& g(x')h(y')-g(x')h(y)-g(x)h(y')+g(x)h(y) \\ 
& \quad\quad = g(x')\big(h(y')-h(y)\big)-g(x)\big(h(y')-h(y)\big) \\
& \quad\quad\geq g(x)\big(h(y')-h(y)\big)-g(x)\big(h(y')-h(y)\big) = 0.
\end{split}
\end{align}
Hence $(x,y)\mapsto g(x)h(y)$ has increasing differences w.r.t.\ any pair $(x,y)\in\mathbb{R}^2$, i.e.\ is supermodular, by \citet[Corollary 1]{Yildiz2010}.~$\square$

The following classical results {on supermodular random vectors} will be instrumental in establishing our main result Theorem \ref{them:MonToSM}.\\

\noindent
\textit{\citet[Definition 2.6]{M97}}\label{def:SuperModRVs}
A random vector $\mathbf{X}=(\mathcal{X}_1,\dots,\mathcal{X}_m)$ is said to be smaller than the random vector $\mathbf{Y}=(\mathcal{Y}_1,\dots,\mathcal{Y}_m)$ in the supermodular ordering, written $\mathbf{X} \leq_{sm} \mathbf{Y}$, if $\mathbb{E}\big(f(\mathbf{X})\big)\leq \mathbb{E}\big(f(\mathbf{Y})\big)$ holds for all the supermodular functions\footnote{see Appendix \ref{apdx:SMFns}.} $f:\mathbb{R}^m\longrightarrow\mathbb{R}$ such that the expectations exist.~$\square$\\

\noindent
\textit{\citet[Definition 3.9.4, page 113]{MS2002}}\label{def:IncSuperModRVs}
A random vector $\mathbf{X}=(\mathcal{X}_1,\dots,\mathcal{X}_m)$ is said to be smaller than the random vector $\mathbf{Y}=(\mathcal{Y}_1,\dots,\mathcal{Y}_m)$ in the increasing supermodular ordering, written $\mathbf{X} \leq_{ism} \mathbf{Y}$, if $\mathbb{E}\big(f(\mathbf{X})\big)\leq \mathbb{E}\big(f(\mathbf{Y})\big)$ holds for all the nondecreasing supermodular functions $f:\mathbb{R}^m\longrightarrow\mathbb{R}$ such that the expectations exist.~$\square$\\

\noindent An equivalent characterization of supermodular vectors is given by\\
\noindent\textit{\citet[Theorems 3.9.11 (i) and (ii), page 118]{MS2002}}\label{th:EquivCaracVecSM} The following statements are equivalent:
\begin{enumerate}
\item $\mathbf{X}\leq_{sm}\mathbf{Y}$,
\item $\mathbf{X}$ and $\mathbf{Y}$ have the same marginals and $\mathbf{X}\leq_{ism}\mathbf{Y}$.~$\square$
\end{enumerate}

\noindent\textit{\citet[Theorems 3.2 (c)]{M97}}\label{th:SuperModRVsProp2}
If $\mathbf{X}$, $\mathbf{Y}$, $\mathbf{Z}$ are random vectors such that any random vectors distributed as $\mathbf{X}$ and $\mathbf{Y}$ conditionally on $\mathbf{Z}=\mathbf{z}$, denoted by $\big[\mathbf{X}|\mathbf{Z}=\mathbf{z}\big]$ and $\big[\mathbf{Y}|\mathbf{Z}=\mathbf{z}\big]$, verify
$\big[\mathbf{X}|\mathbf{Z}=\mathbf{z}\big]\leq_{sm}\big[\mathbf{Y}|\mathbf{Z}=\mathbf{z}\big]$ for all possible values of $\mathbf{z}$, then $\mathbf{X} \leq_{sm} \mathbf{Y}$.~$\square$\bigskip

\noindent\textit{\citet[Definition 2.1]{M97}}\label{def:StopLoss} For $\mathcal{X}$ and $\mathcal{Y}$ in $L^1(\mathbb{Q})$, $\mathcal{X}$ precedes $\mathcal{Y}$ in stop-loss order, written $\mathcal{X} \leq_{sl} \mathcal{Y}$, if
$\mathbb{E}(\mathcal{X} - A)^+ \leq \mathbb{E}(\mathcal{Y} - A)^+$ holds for all real constants $A\geq 0$.~$\square$\\



\noindent\textit{\citet[Theorem 2.2 b)]{BM98}}\label{th:SLequivNondcvxfn} {For any $\mathcal{X}$ and $\mathcal{Y}$, }$\mathcal{X} \leq_{sl} \mathcal{Y}$ if and only if $\mathbb{E}\big(f(\mathcal{X}) \big)\leq \mathbb{E}\big(f(\mathcal{Y}) \big)$ holds for all the nondecreasing convex functions $f:\mathbb{R}\rightarrow\mathbb{R}$ such that the expectations exist (e.g.\ $f=id$).~$\square$
\begin{remark}
The former equivalent condition is also known as {\it the
second-order stochastic dominance}, according to \citet[paragraph 8.1.2, page 267]{MS2012}.
\end{remark}




\section{Elliptical Distributions}\label{s:ellipt}
\noindent\textit{\citet[Definition 6.17, page 196]{McNeilFreyEmbrechts2015}}\label{def:SphericalDef}
A random vector $\mathbf{Z}=(\mathcal{Z}_1,\dots,\mathcal{Z}_n)^\top$ has a spherical distribution in $\mathbb{R}^n$ if, for every orthogonal map $\mathbf{A}\in\mathbb{R}^{n\times n}$ ({i.e. }$\mathbf{A} \mathbf{A}^\top=\mathbf{A}^\top \mathbf{A}=I_n$),
\begin{align}\label{e:SphericalDef}
\mathbf{A}\mathbf{Z}\overset{d}{=}\mathbf{Z}.~\square
\end{align}

\noindent\textit{\citet[Theorem 6.18, page 196]{McNeilFreyEmbrechts2015}}\label{th:SphericalCharac}
The following are equivalent.
\begin{enumerate}
\item[(1)] $\mathbf{Z}$ is spherical in $\mathbb{R}^n$.
\item[(2)] There exists a function $\psi:\mathbb{R}_+\longrightarrow\mathbb{C}$ such that, for all $\mathbf{u}=(u_1,\dots,u_n)^\top\in\mathbb{R}^n$, the characteristic function of $\mathbf{Z}$ is
\begin{align}\label{e:CharcFnSphVb}
\mathbb{E}\left(e^{i\mathbf{u}^\top \mathbf{Z}}\right)=\psi(\mathbf{u}^\top\mathbf{u})=\psi(u_1^2+\cdots+u_n^2).
\end{align}
\item[(3)] For every $\mathbf{u}\in\mathbb{R}^n$, $\mathbf{u}^\top\mathbf{Z}\overset{d}{=}||\mathbf{u}||\mathcal{Z}_1$.~$\square$
\end{enumerate}

\noindent
$\psi$ is {then} called the characteristic generator of $\mathbf{Z}$ and the notation $\mathbf{Z}\sim S_{n}(\psi)$ is used (see \citet*{FKN1990} and \citet*{McNeilFreyEmbrechts2015}). We denote by
$\mathbb{S}^{n-1}:=\big\{\mathbf{s}\in\mathbb{R}^n:\mathbf{s}^\top \mathbf{s}=1\big\}$ the unit sphere in $\mathbb{R}^n$, and by $U_{\mathbb{S}^{n-1}}$, the uniform distribution on $\mathbb{S}^{n-1}$ which has a spherical distribution in $\mathbb{R}^n$ (see \citet*[Example 2.1, page 28 \& Theorem 3.1, page 70]{FKN1990}).\\

\noindent\textit{\citet[Theorem 2.2, page 29]{FKN1990}}\label{pro:GeneratorAndRotation}
A function $\psi$ is a generator of an $n$-dimensional elliptical r.v.\ if and only if it can be written as
\begin{align}\label{e:CharacGenExpr}
\psi(x)=\int_{0}^{\infty}\Omega_n(xr^2)F(dr),
\end{align}
where $F(.)$ is some c.d.f.\ over $\mathbb{R}^+$ and $\Omega_n(\mathbf{u}^\top \mathbf{u})$ is the characteristic generator of a random vector $\mathbf{S}\sim U_{\mathbb{S}^{n-1}}$, namely \citep*[Eqn.~(2)]{Steerneman2005}
\begin{align}
\Omega_n(\mathbf{u}^\top\mathbf{u})
=\mathbb{E}\left(e^{i\mathbf{u}^\top \mathbf{S}}\right)
=\displaystyle\frac{\Gamma(m/2)}{\sqrt{\pi}\Gamma((m-1)/2)}\displaystyle\int_{-1}^1 e^{i \mathbf{u}^\top\mathbf{u} t}\left(1-t^2\right)^{(m-3)/2}dt\sp   
\mathbf{u} \in\mathbb{R}^n. ~\square
\end{align}

{
\noindent\textit{\citet[Theorem 6.21, page 197]{McNeilFreyEmbrechts2015}}\label{th:SphEquivRepr}
$\mathbf{Z}$ has a spherical distribution in $\mathbb{R}^n$ if and only if it has a stochastic representation
\begin{equation}\label{e:SphEquivRepr}
\mathbf{Z}\overset{d}{=}\mathcal{R}\mathbf{S},
\end{equation}
where $\mathbf{S}\sim U_{\mathbb{S}^{n-1}}$ and $\mathcal{R}\geq 0$ is a radial {(i.e. a nonnegative scalar)} r.v.\ independent of $\mathbf{S}$.~$\square$\smallskip
}

\noindent\textit{\citet[Definition 6.25, page 200]{McNeilFreyEmbrechts2015}}\label{def:EllipticalDef}
$\mathbf{X}=(\mathcal{X}_1,\dots,\mathcal{X}_n)$ is said to have an elliptical distribution in $\mathbb{R}^n$ with parameters ${\boldsymbol \mu}$, $\Sigma$, $\psi$, where $\Sigma$ is an $n\times n$ square semi-positive definite matrix, if
\begin{align}\label{e:CharachFECrv}
\mathbb{E}\left(e^{i\mathbf{u}^\top(\mathbf{X}-{\boldsymbol \mu})}\right)=\psi\big(\mathbf{u}^{\top}\Sigma \mathbf{u}\big),\hspace{0.2cm}\mathbf{u}\in\mathbb{R}^n.
\end{align}
We then write $\mathbf{X}\sim E_n({\boldsymbol \mu},\Sigma,\psi)$\footnote{see \citet*{FKN1990} and \citet*{McNeilFreyEmbrechts2015}.}.~$\square$\\

\noindent\textit{\citet[Proposition 6.27, page 200]{McNeilFreyEmbrechts2015}}\label{pro:EllipticEquivCharac}
$\mathbf{X}\sim E_n({\boldsymbol \mu},\Sigma,\psi)$
if and only if there exist $\mathbf{S}$, $\mathcal{R}$ and $\mathbf{A}$ satisfying
$$
\mathbf{X}\overset{d}{=}{\boldsymbol \mu}+\mathcal{R}\mathbf{A}\mathbf{S},
$$
where $\mathbf{S}\sim U_{\mathbb{S}^{k-1}}$, $\mathcal{R}$ is a radial r.v.\ independent of $\mathbf{S}$, and $\mathbf{A}$ 
in $\mathbb{R}^{n\times k}$ satisfies $\mathbf{A}\mathbf{A}^\top=\Sigma$.~$\square$
\begin{remark}\label{rem:FmEllipticToPolar} As outlined in \citet*[Eqn.~(6.41), page 201)]{McNeilFreyEmbrechts2015}, for $\Sigma$ positive definite,
\begin{align}\label{e:FmEllipticToSpheric}
\mathbf{X}\sim E_n({\boldsymbol \mu},\Sigma,\psi)\Longleftrightarrow \Sigma^{-1/2}(\mathbf{X}-{\boldsymbol \mu})\sim S_n(\psi).
\end{align}
Following \citet*[Eqn.~(6.42), page 201)]{McNeilFreyEmbrechts2015}, for an elliptical variate $\mathbf{X}\sim E_n({\boldsymbol \mu},\Sigma,\psi)$, if $\Sigma$ has full rank $n$, then, by \citet[Corollary 6.22, page 198]{McNeilFreyEmbrechts2015} and \eqref{e:FmEllipticToSpheric},
\begin{align}\label{e:FmEllipticToPolar}
\left(\sqrt{(\mathbf{X}-{\boldsymbol \mu})^\top \Sigma^{-1}(\mathbf{X}-{\boldsymbol \mu})},\frac{\Sigma^{-1/2}(\mathbf{X}-{\boldsymbol \mu})}{\sqrt{(\mathbf{X}-{\boldsymbol \mu})^\top \Sigma^{-1}(\mathbf{X}-{\boldsymbol \mu})}}\right)\overset{d}{=}(\mathcal{R},\mathbf{S}).~\square
\end{align}
\end{remark}

{
\noindent\textit{\citet[Theorem 2.18, page 45]{FKN1990}}\label{th:CondElliptical}
Let $\mathbf{Y}\overset{d}{=}{\boldsymbol \mu}+\mathcal{R}\mathbf{A}\mathbf{S}\sim E_{n}({\boldsymbol \mu},\Sigma,\psi)$ (see Section \ref{s:ellipt})
with $\Sigma=\mathbf{A}\mathbf{A}^\top$ positive definite. Let 
\begin{align}\label{e:PartitnParams}
\mathbf{Y}=\begin{pmatrix}\mathbf{Y}^{(1)}\\ \mathbf{Y}^{(2)}\end{pmatrix},\hspace{1cm}{\boldsymbol \mu}=\begin{pmatrix}{\boldsymbol \mu}^{(1)}\\ {\boldsymbol \mu}^{(2)}\end{pmatrix},\hspace{1cm}\Sigma=\begin{pmatrix}\Sigma^{(1,1)} & \Sigma^{(1,2)}\\ \Sigma^{(2,1)} & \Sigma^{(2,2)}\end{pmatrix},
\end{align}
where $\mathbf{Y}^{(1)}$ and ${\boldsymbol \mu}^{(1)}$ are $m\times 1$ vectors and $\Sigma^{(1,1)}$ is an $m\times m$ matrix, for some $0<m<n$. Then
\begin{align}\label{e:Ell}
\begin{split}
\Big(\mathbf{Y}^{(1)}\Big|\mathbf{Y}^{(2)}=\mathbf{y}^{(2)}\Big) & \overset{d}{=} {\boldsymbol \mu}^{(1)}_{|\mathbf{Y}^{(2)}=\mathbf{y}^{(2)}}+\mathcal{R}_{|\mathbf{Y}^{(2)}=\mathbf{y}^{(2)}}\mathbf{A}^{(1,1)}_{|\mathbf{Y}^{(2)}=\mathbf{y}^{(2)}}\mathbf{S}^{(m)} \\
& \sim E_{m}\left({\boldsymbol \mu}^{(1)}_{|\mathbf{Y}^{(2)}=\mathbf{y}^{(2)}},\Sigma^{(1,1)}_{|\mathbf{Y}^{(2)}=\mathbf{y}^{(2)}},\psi_{|\mathbf{Y}^{(2)}=\mathbf{y}^{(2)}}\right),
\end{split}
\end{align}
where
\begin{align}\label{e:EllParamsCond}
\left\{
\begin{array}{l}
     {\boldsymbol \mu}^{(1)}_{|\mathbf{Y}^{(2)}=\mathbf{y}^{(2)}} = {\boldsymbol \mu}^{(1)}+\Sigma^{(1,2)}\big(\Sigma^{(2,2)}\big)^{-1}\left(\mathbf{y}^{(2)}-{\boldsymbol \mu}^{(2)}\right), \\
     \\
     \Sigma^{(1,1)}_{|\mathbf{Y}^{(2)}=\mathbf{x}^{(2)}} =
     \Sigma^{(1,1)}-\Sigma^{(1,2)}\big(\Sigma^{(2,2)}\big)^{-1}\Sigma^{(2,1)}= \mathbf{A}^{(1,1)}_{|\mathbf{Y}^{(2)}=\mathbf{y}^{(2)}}\left(\mathbf{A}^{(1,1)}_{|\mathbf{Y}^{(2)}=\mathbf{y}^{(2)}}\right)^\top,\\
     \\
     \mathbf{S}^{(m)} \sim U_{\mathbb{S}^{m-1}},\\
     \\
     \mathcal{R}_{|\mathbf{Y}^{(2)}=\mathbf{y}^{(2)}} \overset{d}{=} \left(\left(\mathcal{R}^2-q\left(\mathbf{y}^{(2)}\right)\right)^{1/2}\bigg|\mathbf{y}^{(2)}=\mathbf{y}^{(2)}\right)\,\mbox{ and } \mathcal{R}_{|\mathbf{Y}^{(2)}=\mathbf{y}^{(2)}} \mbox{ is independent} \\
\mbox{of } \mathbf{S}^{(m)}, \\
\\
     q\left(\mathbf{y}^{(2)}\right) = \left(\mathbf{y}^{(2)}-{\boldsymbol \mu}^{(2)}\right)^\top \big(\Sigma^{(2,2)}\big)^{-1}\left(\mathbf{y}^{(2)}-{\boldsymbol \mu}^{(2)}\right), \\ \\
\text{$\psi_{|\mathbf{Y}^{(2)}=\mathbf{y}^{(2)}}$ is of the form \eqref{e:CharacGenExpr} for $n=m$, $F$ given as the c.d.f. of $\mathcal{R}_{|\mathbf{Y}^{(2)}=\mathbf{y}^{(2)}}$ and} \\
\text{$\mathbf{S}$ given as $\mathbf{S}^{(m)}$}.~\square
\end{array}
\right.
\end{align}
}

\noindent\textit{\citet*[Corollary 2.3]{BS88}}\label{cor:EllipticSMpropCor}
Let $\mathbf{X}\sim E_n({\boldsymbol \mu},\Sigma,\psi)$ and $h:\mathbb{R}^n\rightarrow\mathbb{R}$ be a supermodular, bounded and right-continuous function. Then $\mathbb{E}\big(h(\mathbf{X})\big)$ is nondecreasing in the off-diagonal elements of $\Sigma$.~$\square$

\noindent
The extension of \citet*[Corollary 2.3]{BS88} to all supermodular functions follows from \citet*[Theorem 3.3 and Theorem 3.4]{MS2000}. Hence, if $\mathbf{X}$, $\mathbf{X}'$ are defined as per \eqref{e:ModelElliptic}, then $\mathbf{X}\leq_{sm}\mathbf{X}'$. As a direct consequence outlined in \citet*{SS97} after their Definition 2.1, ``it follows that the family of multivariate normal distributions (more generally, the family of elliptically contoured distributions) is increasing in the supermodular stochastic order as the correlations increase.''


\subsection{Preserving Elliptical Distributions with Conditioning}\label{ss:EllipticCase}

\noindent Let $\boldsymbol \mu=(\mu_1,\dots,\mu_m)$ and $\boldsymbol\Gamma^\mathbf{X}=\Big(\Gamma_{ij}\Big)_{1\leq i,j\leq m}$ be the mean vector and the covariance matrix of $(\mathcal{X}_1,\dots,\mathcal{X}_m)$ under $\mathbb{Q}$.
We use similar notations ${\boldsymbol \mu}'$ and $\boldsymbol\Gamma^{\mathbf{X}'}$ regarding $(\mathcal{X}'_1,\dots,\mathcal{X}'_m)$.




\begin{lemma}\label{lem:SpecCaseElliptic}
Under the elliptical model setup \eqref{e:ModelElliptic} for $(\mathcal{X}_0,\mathbf{X})$ and $(\mathcal{X}_0,\mathbf{X}')$, we have
\begin{align}\label{e:CondX0Distri}
\begin{split}
& [\mathbf{X}|\mathcal{X}_0]\sim E_m\left({\boldsymbol \mu}_{|0},\boldsymbol\Gamma_{|0},\psi_{|0}\right), \\ 
& [\mathbf{X}'|\mathcal{X}_0]\sim E_m\left({\boldsymbol \mu}'_{|0},\boldsymbol\Gamma'_{|0},\psi'_{|0}\right),
\end{split}
\end{align}
with
\begin{align}\label{:SpecCaseEllipticMeans}
\psi_{|0}=\psi'_{|0},\,\,
{\boldsymbol \mu}_{|0}={\boldsymbol \mu}'_{|0}
\end{align}
and, for any $i,j\in 1\, ..\, m$,
\begin{align}\label{e:SpecCaseElliptic}
    \Gamma_{|0} \leq \Gamma'_{|0},
\end{align}
where $\leq$ is meant componentwise.
\end{lemma}
\proof
Applying \eqref{e:EllParamsCond} to $\mathbf{Y}^{(1)}=\mathbf{X}$ and $\mathbf{Y}^{(2)}=\mathcal{X}_0$ yields \eqref{e:CondX0Distri} with
\begin{align}\label{e:CondParamsX1n}
\begin{array}{l}
    {\boldsymbol \mu}_{|0}={\boldsymbol \mu}+\displaystyle\frac{1}{\Gamma_{00}}\Big(\Gamma_{01}{,}\dots,\Gamma_{0m}\Big)^\top (\mathcal{X}_0-\mu_{0}), \\
     \boldsymbol\Gamma_{|0}=
     \boldsymbol\Gamma^{\mathbf{X}}-\displaystyle\frac{1}{\Gamma_{00}}{\Big(\Gamma_{01},\dots,\Gamma_{0m}\Big)^\top\Big(\Gamma_{01},\dots,\Gamma_{0m}\Big)}
\end{array}
\end{align}
and, using \eqref{e:CondParamsX1n},
\begin{align}\label{e:CondParamsX1n_prime}
\begin{array}{l}
     {\boldsymbol \mu}'_{|0}={\boldsymbol \mu}+\displaystyle\frac{1}{\Gamma_{00}}\Big(\Gamma_{01}{,}\dots,\Gamma_{0m}\Big)^\top (\mathcal{X}_0-\mu_{0})={\boldsymbol \mu}_{|0} \\
     \boldsymbol\Gamma'_{|0}=
     \boldsymbol\Gamma^{\mathbf{X}'}-\displaystyle\frac{1}{\Gamma_{00}}{\Big(\Gamma_{01},\dots,\Gamma_{0m}\Big)^\top\Big(\Gamma_{01},\dots,\Gamma_{0m}\Big)}\geq \boldsymbol\Gamma_{|0}.
\end{array}
\end{align}
where $\geq$ is meant componentwise.\smallskip

It remains to show that $\psi_{|0}=\psi'_{|0}$. For all $x_0\in\mathbb{R}$, by \citet[Theorem 2.18, page 45]{FKN1990}, which includes \eqref{e:EllParamsCond}, the radius $\mathcal{R}_{|\mathcal{X}_0=x_0}$ of $[\mathbf{X}|\mathcal{X}_0=x_0]$ is distributed like $\left(\mathcal{R}^2 - \displaystyle\frac{1}{\boldsymbol\Gamma^{00}}(x_0-\mu_{\mathcal{X}_0})^2\right)^{1/2}$, where $\mathcal{R}:=\|\mathbf{Z}\|$, and so is the radius $\mathcal{R}'_{|\mathcal{X}_0=x_0}$ of $[\mathbf{X}'|\mathcal{X}_0=x_0]$ (as both $(\mathcal{X}_0,\mathbf{X})$ and $(\mathcal{X}_0,\mathbf{X}')$ are defined based on the spherically distributed vector $\mathbf{Z}$). Thus $\mathcal{R}_{|\mathcal{X}_0}\overset{d}{=}\mathcal{R}'_{|\mathcal{X}_0}$,
with common c.d.f.\ denoted by $F_{\mathcal{R}_{|\mathcal{X}_0}}$. The corresponding conditional characteristic generator common to $\left[\mathbf{X}|\mathcal{X}_0\right]$ and $\left[\mathbf{X}'|\mathcal{X}_0\right]$ is given by $\psi_{|\mathcal{X}_0}(x)=\displaystyle\int_{0}^{\infty} \Omega_{m}(xr^2)F_{\mathcal{R}_{|\mathcal{X}_0}}(dr)$, where $\Omega_{m}(u^\top u)$ is the characteristic function of a 
{random variable that is uniformly distributed on the unit sphere of $\mathbb{R}^{m}$}.~$\square$

\begin{proposition}\label{p:ConclResFmLemmas}
Under the assumptions of Lemma \ref{lem:SpecCaseElliptic}, we have:
\begin{align}\label{e:SMpropPresSurvMs}
[\mathbf{X}|\mathcal{X}_0]\leq_{sm} [\mathbf{X}'|\mathcal{X}_0].
\end{align}
\end{proposition}
\proof
By Lemma \ref{lem:SpecCaseElliptic}, conditionally on $\mathcal{X}_0$, 
$\mathbf{X}$ and $\mathbf{X}'$ have the same elliptical distribution  under $\mathbb{Q}$, except for their covariance matrix coefficients that verify \eqref{e:SpecCaseElliptic}. \citet*[Corollary 2.3]{BS88} recalled in Section \ref{s:ellipt} (here applied under $\mathbb{Q}$) then yields the result.~$\square$


\section{Risk Measures}\label{s:RMordPpties}
The following definition of a risk measure relaxes the definition of a convex risk measure in \citet*[Introduction (A3)]{Shapiro2013} by not requiring the translation-equivariance property.
\begin{definition}\label{d:ConvexRiskMeas}
Let $\mathfrak{X}$ be a closed linear subspace and sublattice of $L^1(\mathbb{Q})$, that includes the
constants. 
A risk measure on $\mathcal{X}$ is a function $\rho:\mathfrak{X}\rightarrow\mathbb{R}$ satisfying, for $\mathcal{X},\mathcal{Y}\in\mathfrak{X}$:
\begin{enumerate}[label=\Alph*]
    \item properness: $\rho\left(\mathcal{X}\right)>-\infty$ and $\mathrm{dom}\,\rho:=\left\{\mathcal{X}\in\mathfrak{X};\rho\left(\mathcal{X}\right)<+\infty\right\}\neq\varnothing$;
    \item law-invariance: if $\mathcal{X}\stackrel{d}{=}\mathcal{Y}$ then $\rho\left(\mathcal{X}\right)=\rho\left(\mathcal{Y}\right)$;
    \item monotonicity: if $\mathcal{X}\leq\mathcal{Y}$ then $\rho\left(\mathcal{X}\right)\leq\rho\left(\mathcal{Y}\right)$;
        \item convexity: $\rho\left(\lambda\mathcal{X}+(1-\lambda)\mathcal{Y}\right)\leq \lambda\rho\left(\mathcal{X}\right)+(1-\lambda)\rho\left(\mathcal{Y}\right)$, $\lambda\in(0,1)$.~$\square$
\end{enumerate}
\end{definition}


Certain risk measures are also translation-equivariant (i.e. $\rho\left(\mathcal{X}+c\right)=\rho\left(\mathcal{X}\right)+c$, $c\in\mathbb{R}$ see \citep[Definition 2.2 (i), page 29]{PflugWerner2007}), such as:
\begin{definition}\label{d:ESdef}
The expected shortfall (ES)\footnote{see \citet[Definition 2.6]{AT2002}.} at the confidence level (quantile) $\alpha\in\left(\frac{1}{2},1\right)$ of a loss $\mathcal{X}\in\mathfrak{X}=L^1(\mathbb{Q})$ is $\mathbb{ES}_{\alpha}(\mathcal{X})=(1-\alpha)^{-1}\Big(\mathbb{E}\big(\mathcal{X}\mathds{1}_{\{\mathcal{X}\ge
{\mathbb{V}\mathrm{a}\mathbb{R}_\alpha}(\mathcal{X})\}}\big)+\mathbb{V}\mathrm{a}\mathbb{R}_\alpha(\mathcal{X})\big(\mathbb{Q}(\mathcal{X}<\mathbb{V}\mathrm{a}\mathbb{R}_\alpha(\mathcal{X}))-\alpha\big)\Big)$, with $\mathbb{V}\mathrm{a}\mathbb{R}_\alpha(\mathcal{X})=\inf\big\{x\in\mathbb{R}:\mathbb{Q}(\mathcal{X}\leq x)>\alpha\big\}.$~$\square$
\end{definition}
\begin{remark}\label{rem:rhoAsES}
The domain of $\mathbb{ES}^0_{\alpha}$ is all $L^1(\mathbb{Q}^0)$, thus {$\mathbb{ES}^0_{\alpha}$} is in particular proper. \citet*[Proposition 3.1]{AT2002} outlines that $\mathbb{ES}^0_{\alpha}$ is subbaditive and positively homogeneous (therefore convex) as well as monotonous (in our case where we consider loss variables, for $\mathcal{X}\leq 0$ we have $\rho(\mathcal{X}) \leq 0$ and using subadditivity we obtain for $\mathcal{X}\leq\mathcal{Y}$ that $\rho(\mathcal{X}) =\rho(\mathcal{X}-\mathcal{Y}+\mathcal{Y})\leq\rho(\mathcal{X}-\mathcal{Y})+\rho(\mathcal{Y}) \leq\rho(\mathcal{Y}) $). Hence $\mathbb{ES}^0_{\alpha}$ is a risk measure in the sense Definition \ref{s:RMordPpties}.
\end{remark}

We conclude this appendix by a detailed proof of \citet*[Theorem 4.4]{BM05}.
The proof relies on the definition of stochastic order:\smallskip\smallskip

\noindent\textit{\citet*[Definition 2.1 a)]{BM98}}\label{def:StochasticOrder} For $\mathcal{X}$ and $\mathcal{Y}$ with respective c.d.f.\ $F_{\mathcal{X}}$ and $F_{\mathcal{Y}}$, $\mathcal{X}$ precedes $\mathcal{Y}$ in stochastic order, written $\mathcal{X} \leq_{st} \mathcal{Y}$, if
$F_{\mathcal{X}}(x) \geq F_{\mathcal{Y}}(x)$, $x\in\mathbb{R}$.~$\square$\\

\begin{remark}
The former order definition is also known as {\it the
first-order stochastic dominance}, according to \citet[paragraph 8.1.2, page 266]{MS2012}.
\end{remark}

\noindent\textit{\citet*[Theorem 4.4]{BM05}}\label{them:SLtoRM} Assuming $\rho$ a risk measure as per Definition \ref{d:ConvexRiskMeas}, then{, for any $\mathcal{X}$ and $\mathcal{Y}$ in $\mathfrak{X}$}, $\mathcal{X} \leq_{sl} \mathcal{Y}$ implies $\rho\left(\mathcal{X}\right) \leq\rho\left(\mathcal{Y}\right) $.~$\square$\smallskip

 
\proof
A succinct proof of this result can be found in \citet*{BM05} prior to its statement, but, for completeness, we give a more detailed proof hereafter.
If $\mathcal{X}\leq_{sl} \mathcal{Y}$, then, 
by \citet*[Theorem 1.5.14, page 22]{MS2002}, there exists a r.v.\ $\mathcal{Z}\in\mathfrak{X}$ s.t.\ $\mathcal{X}\leq_{st}\mathcal{Z}\leq_{cx}\mathcal{Y}$, where $\leq_{cx}$ is the convex order. 
By \citet*[Theorem 1.2.4, page 3]{MS2002}, there exist r.v. $\mathcal{X}'$ and $\mathcal{Z}'$ on a modified probability space $(\Omega',\mathcal{A}',\mathbb{Q}')$, with same respective laws as $\mathcal{X}$ and $\mathcal{Z}$, such that $\mathcal{X}'\leq \mathcal{Z}'$ holds with certainty i.e., $\forall \omega\in\Omega'$, $\mathcal{X}'(\omega)\leq \mathcal{Z}'(\omega)$ where $\leq$ is the partial order\footnote{see \citet*[Remark 1.2.5, page 3]{MS2002}.} on $\mathbb{R}^m$ if $\mathcal{X}$ and $\mathcal{Z}$ take value in $\mathbb{R}^m$. The law-invariance and monotonicity of $\rho$ yield
\begin{align}\label{e:RiskStOrd}
\rho\left(\mathcal{X}\right)=\rho\left(\mathcal{X}'\right)\leq\rho\left(\mathcal{Z}'\right)=\rho\left(\mathcal{Z}\right).
\end{align}
From \citet*[Theorem 4.3]{BM05}, which requires the convexity and law-invariance of $\rho$, we also have $\rho\left(\mathcal{Z}\right)\leq\rho\left(\mathcal{Y}\right)$, hence $\rho\left(\mathcal{X}\right)\leq \rho\left(\mathcal{Y}\right)$.~$\square$

\begin{remark}\label{rem:FatouNotReq}
Note that the original statement postulates, instead of the law-invariance property, that the risk measure $\rho$ has the Fatou property, that is, if $\mathcal{X},\mathcal{X}_1,\mathcal{X}_2,\dots$ are integrable random variables with $\mathcal{X}_k\xrightarrow[]{L^1} \mathcal{X}$, then $\rho\left(\mathcal{X}\right) \leq\liminf_{k\rightarrow\infty}\rho\left(\mathcal{X}_k\right)$. We recall from \citet[page 832]{Kallenberg2021} that a closed linear subspace of a Banach space is a Banach space and from \citet[Definition 1.2.1 i), page 12]{MeyerNieberg1991} that a sublattice of a lattice with the same meet and join operations $\wedge$ and $\vee$ is again a lattice. Therefore, if $\mathfrak{X}$ is a Banach lattice (i.e.\ an order lattice that is a complete normed vector space, e.g.\ $L^p$ space with $p\geq 1$) and $\rho:\mathfrak{X}\rightarrow\mathbb{R}$ is proper, monotonous and convex, then $\rho$ is continuous on the interior of its domain \citep[Proposition 1]{RS2006}, thus has the Fatou property on the interior of its domain. Hence the Fatou property requirement is automatically satisfied by $\rho$ as long as it is defined on a Banach lattice. This is the case for both expectation and expected shortfall defined on any sublattice and linear subspace of $L^1$.
\smallskip
\end{remark}

\section*{Acknowledgements}

We thank Mekonnen Tadese, postdoctoral researcher at Université Paris Cité / LPSM, for useful references and discussions.\smallskip

\noindent We thank the two anonymous reviewers for their meticulous in-depth review and valuable suggestions that helped improve the quality of the paper.

\section*{Funding}

This research benefited from the support of the \textit{Chair Stress Test, RISK Management and Financial Steering}, led by the French \'Ecole polytechnique and its Foundation and sponsored by BNP Paribas.

\end{document}